\documentclass[a4paper, twocolumn, accepted=2024-06-30]{quantumarticle}
\pdfoutput=1

\usepackage{preamble_lat}
\hypersetup{
pdftitle={Good Gottesman-Kitaev-Preskill codes from the NTRU cryptosystem},
pdfsubject={Good Gottesman-Kitaev-Preskill codes from the NTRU cryptosystem},
pdfauthor={Jonathan Conrad, Jens Eisert, Jean-Pierre Seifert},
pdfkeywords={Quantum error correction, lattice theory, continuous variable codes, quantum cryptography}
}

\usepackage{tabularx}

\begin{document}


\title{Good Gottesman-Kitaev-Preskill codes from the NTRU cryptosystem} 

\author{Jonathan Conrad}\orcid{0000-0001-6120-9930}
\email{j.conrad1005@gmail.com}
\affiliation{Dahlem Center for Complex Quantum Systems, Physics Department, Freie
Universit{\"a}t Berlin, Arnimallee 14, 14195 Berlin, Germany}
\affiliation{Helmholtz-Zentrum Berlin f{\"u}r Materialien und Energie, Hahn-Meitner-Platz 1, 14109
Berlin, Germany}

\author{Jens Eisert}\orcid{0000-0003-3033-1292}
\affiliation{Dahlem Center for Complex Quantum Systems, Physics Department, Freie
Universit{\"a}t Berlin, Arnimallee 14, 14195 Berlin, Germany}
\affiliation{Helmholtz-Zentrum Berlin f{\"u}r Materialien und Energie, Hahn-Meitner-Platz 1, 
14109 Berlin, Germany}
\affiliation{Fraunhofer Heinrich Hertz Institute, Einsteinufer 37, 10587 Berlin, Germany}

\author{Jean Pierre Seifert}\orcid{0000-0002-5372-4825}
\affiliation{Electrical Engineering and Computer Science Department,
Technische Universit{\"a}t Berlin, Stra{\ss}e des 17. Juni 135,
10587 Berlin, Germany}
\affiliation{Fraunhofer Institute for Secure Information Technology,
Rheinstra{\ss}e 75,
64295 Darmstadt, Germany}

\date{10.4.2023}

\maketitle
\begin{abstract}
We introduce a new class of random Gottesman-Kitaev-Preskill (GKP) codes derived from the cryptanalysis of the so-called NTRU cryptosystem.  The derived codes are \textit{good} in that they exhibit constant rate and average distance scaling $\Delta \propto \sqrt{n}$ with high probability,  where $n$ is the number of bosonic modes,  which is a distance scaling equivalent to that of a GKP code obtained by concatenating single mode GKP codes into a qubit-quantum error correcting code with linear distance. 
The derived class of NTRU-GKP codes has the additional property that \textit{decoding} for a stochastic displacement noise model is equivalent to \textit{decrypting} the NTRU cryptosystem,  such that every random instance of the code naturally comes with an efficient decoder.  This construction highlights how the GKP code bridges aspects of classical error correction,  quantum error correction as well as post-quantum cryptography.  We underscore this connection by discussing the computational hardness of decoding GKP codes and propose,  as a new application,  a simple public key quantum communication protocol with security inherited from the NTRU cryptosystem.
\end{abstract}

\section{Introduction}
In recent years,  notions of bosonic quantum-error correction with the \emph{Gottesman-Kitaev-Preskill} (GKP) code \cite{GKP} have seen a rapid increase of interest
both in theory and in experiment,
primarily due to the perspective of 
them contributing to a
viable route towards large scale quantum computing using 
integrated \emph{photonic}
\cite{Bourassa_2021, bartolucci2021fusionbased} 
and \emph{superconducting platforms} \cite{GrimsmoPuri}. 
Such codes have highly attractive features for
systems in which quantum information is encoded in 
continuous variable degrees of freedom and are 
specifically suitable to accommodate photon 
loss \cite{Albert_2018, Noh_Capacity, Terhal_2020, HastrupLoss}. 
While much research has been dedicated to obtain effective qubits from single-mode systems that are to be integrated into larger qubit-based networks \cite{toricGKP, Noh_2020}, 
this approach is arguably only scratching the surface of possibilities offered by the GKP code within its more general,  lattice theoretic perspective \cite{Harrington_Thesis,  Conrad_2022,  Royer_2022,  Schmidt_2022}. 

To corroborate this claim,  in this work,
we construct random \textit{good} GKP codes derived from a cryptographic attack on the NTRU cryptosystem \cite{Hoffstein_Pipher_Silverman_1998}.  We define and discuss a goodness property for GKP codes in the lattice theoretic framework analogous to the notion of goodness in conventional quantum- 
and classical error correcting codes as a stepping stone towards scalable codes. 
We investigate the decoding problem of this class of codes and show how the native \textit{decryption} routine of the NTRU cryptosystems with access to its secret key can serve as decoder for the corresponding GKP code.

We investigate the complexity of decoding general GKP codes and highlight how our \textit{NTRU-GKP} codes can be viewed as a \textit{trapdoor decodable quantum error correcting code},  where decoding of the associated quantum error correcting code is expected to by computationally hard in general but becomes significantly easier when supplemented with additional (secret) information about the structure of the code.  We consider this a first 
step towards cryptographic protocols built on the decoding problem of GKP codes which we believe can have wide application for secure quantum communication and cloud-based quantum computing and hope that this article stimulates interest in this new direction of research.

This article is structured as follows.  In Section \ref{sec:GKP}, we introduce basic principles of the GKP code,  describe relevant aspects of the general decoding problem for the GKP code and define \textit{goodness} of a GKP code family.  In Section
\ref{sec:codes}, we discuss a selection of notable examples of GKP codes and summarize their properties.  The NTRU cryptosystem and GKP codes built on this are discussed in Section \ref{sec:NTRUGKP} where we provide evidence that they form a family of \textit{good} randomized GKP codes.

The goodness property of GKP codes is established through proposition \ref{prop_1} and conjecture \ref{conj:random_h}, \ref{conj:SS}. Proposition \ref{prop_1} establishes that NTRU-GKP codes built on 
schemes to sample the public key prescribing the NTRU lattice according to the 
procedure originally proposed in ref.\ \cite{Hoffstein_Pipher_Silverman_1998},  often referred to as NTRU-HPS, are good with overwhelming probability. This proposition foots on a proof provided by Bi and Qi in 
ref.\ \cite{Jingguo_Qi} and establishes goodness for GKP codes built on a specific distribution of ($q-$symplectic) matrices and lattices.  We observe 
numerically that random public keys for the NTRU cryptosystem well satisfy the good scaling property which is summarized in conjecture \ref{conj:random_h}. Finally,  a variation of the NTRU cryptosystem has been proposed by Stehle and Steinfeld \cite{StehleSteinfeld}, who show that their procedure yields pseudo-random public keys for the NTRU cryptosystem. We numerically confirm the expected goodness of NTRU-GKP codes drawn according to their procedure which is summarized in conjecture \ref{conj:SS}. An additional rigorously minded   perspective is provided in the appendix, where with proposition \ref{prop:randsymH} we establish average-case goodness for a class of symplectic lattices slightly larger than that produced by the NTRU-cryptosystem.  This proof is constructive.
We discuss the decoding problem for the NTRU-GKP codes in Section \ref{sec:decodingNTRU} and discuss a quantum \emph{public key cryptosystem} (PKC) designed around the NTRU-GKP codes. 
Finally,  we conclude and provide outlook in Section \ref{sec:conclusion}.

\section{The GKP code}\label{sec:GKP}
We build our framework on our recent exposition \cite{Conrad_2022} which we refer to for an in-depth discussion; see also the original work \cite{GKP}  as well as refs.~\cite{Harrington_Thesis,  HarringtonPreskill,  Royer_2022, Schmidt_2022}. The \emph{GKP code} \cite{GKP} is a \emph{stabilizer code}  acting on the Hilbert space of $n$ bosonic modes,  where stabilizers are given by displacement operators
\begin{align}
D\lr{\bs{\xi}} = \exp\left\{-i \sqrt{2\pi} \bs{\xi}^T J \bs{\hat{x}}\right\}, \label{eq:displConvJ}
\end{align}
where \begin{align}
J_{2n}=\begin{pmatrix}0 & 1 \\ -1 & 0\end{pmatrix} \otimes I_n= \begin{pmatrix}
0 &  I_n \\
-I_n  & 0 \end{pmatrix}
\end{align} 
is the symplectic form,  $\bs{\xi}\in\mathbb{R}^{2n}$, 
and $\bs{\hat{x}}=\lr{\hat{q}_1,\hat{q}_2,\dots ,\hat{p}_{n-1} ,\hat{p}_n }^T$ is the generalized quadrature operator.   Its stabilizer group is specified by fixing $2n$ linearly independent vectors $\bs{\xi}_i,\,i=1, \dots, 2n$,
\begin{equation}
\CS=\big\langle D\lr{\bs{\xi}_1}\hdots D\lr{\bs{\xi}_{2n}}   \big\rangle=\lrc{e^{i\phi_M\lr{\bs{\xi}}}D\lr{\bs{\xi}},\; \bs{\xi}\in \CL},  \label{eq:GKPdef}
\end{equation}
where $M=\lr{\bs{\xi}_1 , \hdots ,\bs{\xi}_{2n} }^T$ is a generator for the lattice $\CL= \mathbb{Z}^{2n}M$ \footnote{For clarity of presentation,  we sometimes assume row- 
vs.~column vector conventions to be clear from context.} 
and we have
\begin{equation}
\phi_M\lr{\bs{\xi}}=\pi \bs{a}^T A_{\lowertriangle} \bs{a}, \; \bs{a}^T=\bs{\xi}^TM^{-1},
\end{equation}
to denote the phase-sector when we have chosen $M$ to be the basis for which each associated displacement operator is fixed to eigenvalue $+1$ by eq.~\eqref{eq:GKPdef}.  $A=MJM^T$ denotes the symplectic Gram matrix and $A_{\lowertriangle}$ its left lower triangle.

While the stabilizer group $\CS \sim \CL$ is isomorphic to the lattice $\CL$,  its centralizer 
 within the displacements  $\CC\lr{\CS} \sim \CL^{\perp}$ -- i.e., the set of displacement operators that commute with every element in $\CS$ -- is isomorphic to its symplectic dual lattice $\CL^{\perp}:=\lrc{\bs{x} \in \mathbb{R}^{2n},\,  \bs{x}^TJ\bs{\xi} \in \mathbb{Z}\, \forall \bs{x} \in  \CL}$.  $\CS$ is Abelian iff $\CS \subseteq \CC\lr{\CS}$ which is equivalent to $\CL \subseteq \CL^{\perp}$ and $A$ being integer.  
Compactly, GKP codes are represented by \textit{weakly symplectically self-dual} lattices $\CL \subseteq \CL^{\perp}$ where the elements in the symplectic dual quotient $\CL^{\perp}/\CL$ label the group of logical Pauli-operations of the code encoding $D$ dimensions,  which has size $|\det\lr{A}|=D^2$.  When we encode collections of qubits,  $k=\log_2\lr{D}=\log_2\lr{|\det\lr{M}|}=\log_2\lr{|\det\lr{\CL}|}$ denotes the number of encoded logical qubits,  which grows logarithmically with the determinant of the stabilizer lattice.

\paragraph{Code distance.}
The (Euclidean) code \emph{distance} of the GKP code is defined as $\Delta=\min_{\bs{x} \in \CL^{\perp}\setminus \CL} \|x\|$,  the length of the shortest vector in $\CL^{\perp}$ not in $\CL$.  The distance as defined here is a meaningful indicator for the code performance in a stochastic displacement error model if the probability that a certain displacement is realized scales inversely with its length.

The distance is also the smallest power of $q$ in the complex polynomial
\begin{equation}
Q_{\CL}\lr{z}=\Theta_{\CL^{\perp}}\lr{z}-\Theta_{\CL}\lr{z}=N_{\Delta^2}q^{\Delta^2}+ O\lr{q^{\Delta^2}},
\end{equation}
where $q=e^{i\pi z}$,  $z \in \mathcal{H}=\{z \in \mathbb{C},\,  \Im z >0\}$ and 
\begin{equation}
\Theta_{\CL}\lr{z}=\sum_{\bs{x}\in \CL} q^{\| \bs{x}\|^2}
\end{equation}
is the \emph{theta function} of the lattice $\CL$ and $N_{\Delta^2}$ in $Q_{\CL}\lr{z}$ counts the number of lattice points in $\CL^{\perp}\setminus \CL$ of squared length $\Delta^2$.  It is expected that the relative scaling of $N_{\Delta^2}$ with $\Delta^2$ significantly impacts the existence and scale of the threshold of a GKP code family and is responsible for the entropic contribution to the decoding problem \cite{toricGKP, Dennis_2002}. We comment on this further in Appendix \ref{appendix:threshold}. 
The theta function and distance are by construction symmetric under orthogonal transformations $O\in \mathcal{O}\lr{2n}$ of the lattice $\CL\mapsto O\CL$,  
 while $N_{\Delta^2}$ scales with the number of orthogonal automorphisms of the lattice.  

\paragraph{Decoding GKP codes via the closest vector problem.}

Upon measuring the stabilizers on displacement error $D\lr{\bs{e}}$ and obtaining the syndromes $\bs{s}=MJ\bs{e} \mod 1$,  one strategy is to correct back to the code space by applying a displacement in phase space with $\bs{\eta}=\lr{MJ}^{-1}\bs{s}$.  Since the choice of the generator $M$ is ambiguous,  it is generally necessary to append this initial correction by a logical post-correction to minimize the probabilty of imposing a logical error.  Algorithms finding the correction that minimizes logical errors given the syndromes are called \emph{decoders} in this context. 
Assuming a Gaussian displacement error model with variance $\overline{\sigma}^2$ \footnote{We use the overline to denote the rescaled variance that for shifts implemented by the non-standard choice of displacement operators in 
eq.~\eqref{eq:displConvJ}.  The physical variance $\sigma^2$ is related to it by $\sigma^2=2\pi \overline{\sigma}^2$.},  one derives the optimal post-correction \cite{Conrad_2022} to be applied after an initial correction $\bs{\eta}$ to be given by \emph{maximum likelihood decoding} (\texttt{MLD}) 

\begin{equation}
\overline{\bs{\xi}}^{\perp}=\argmax_{\bs{\xi}^{\perp}\in \CL^{\perp}/\CL} \Theta_{\bs{\eta}+\bs{\xi}^{\perp}+\CL}\lr{\frac{i}{2\pi\overline{\sigma}^2}}. \label{eq:MLD}
\end{equation}
For small error rates $\overline{\sigma}\rightarrow 0$,  the most likely coset as computed in \texttt{MLD} is given by the most likely individual error consistent with the syndrome.  We refer to decoding based on the most likely individual error consistent with the syndrome as \textit{minimum energy decoding} (MED) \cite{toricGKP},  to which the solution is presented by the \emph{closest vector problem} (\texttt{CVP}),  $\overline{\bs{\xi}}^{\perp}=\CVP\lr{\bs{\eta}, \,  \CL^{\perp}}$ (see ref.~\cite{Conrad_2022}),  that is,  in this limit \texttt{MLD} reduces to \texttt{CVP}.

\paragraph{Bounded distance decoding (\texttt{BDD}).}
By nature of the Gaussian error model,  it is unlikely to sample an error larger than $\|\bs{e}\|>\sqrt{2 n\overline{\sigma}^2}$ \cite{HarringtonPreskill, Harrington_Thesis}.  Hence,  it is reasonable to restrict the decoding problem to \textit{bounded-distance-decoding} ($\BDD_{\epsilon}\lr{\bs{\eta},  \CL^{\perp}}$),  which is \texttt{CVP} with an additional promise that $\text{dist}\lr{\bs{\eta},  \CL^{\perp}}\leq \epsilon$.  Given that typical Gaussian errors will be overwhelmingly of length $\|\bs{e}\| \leq \sqrt{2 n\overline{\sigma}^2}$ we expect to decode successfully by solving $\BDD_{\epsilon}\lr{\bs{\eta},  \CL^{\perp}}$ with $\epsilon=\sqrt{2 n\overline{\sigma}^2}$.
Provided a lattice basis $M$ and its Gram-Schmidt orthogonalization $\tilde{M}=(\bs{\tilde{\xi}}_1,\hdots, \bs{\tilde{\xi}}_{2n})^T$ it is known that Babai's nearest plane algorithm solves \texttt{BDD} when $\epsilon < \min_i\|\bs{\tilde{\xi}}_i \|/2=:\|\tilde{M}\|/2$ \cite{babai, CryptoLectureNotes},  that is when we are in possession of a lattice basis such that its Gram-Schmidt reduced vectors are sufficiently long
$\|\tilde{M}\|\geq \sqrt{8 n \overline{\sigma}^2}$.

\newtcolorbox{mybox}{colback=white, colframe=red!45!blue}

\begin{figure}
\footnotesize
\begin{algorithmic}[h]
\Function{$\mathtt{NearestPlane}$}{$B,  \bs{t}$}
\If{$len\lr{B}==0$}
\State return $0^{len\lr{\bs{t}}}$
\Else
\State set $B^* = \mathtt{GramSchmidt}\lr{B}$
\State set $c=\lfloor \bs{t}^TB^*_{-1} /\|B^*_{-1}\|^2  \rceil $
\State return $cB_{-1}+\mathtt{NearestPlane}\lr{B_{:-1},  \bs{t}-c B_{-1}}$
\EndIf
\EndFunction
\end{algorithmic}
\caption{Babai's nearest plane algorithm \cite{babai}.  }
\end{figure}

When considering code families with growing dimension $2n$,  errors up to the typical length $\sqrt{2n\overline{\sigma}^2}$ are correctible via CVP decoding only if $\Delta=\Omega\lr{\sqrt{n}}$.  The scaling $\Delta\propto\sqrt{n}$ is the distance scaling of a GKP code obtained by concatenating fixed single-mode GKP codes with a qubit quantum error correcting code with linear distance $d\propto n$.  Such scaling in error correction performance is predicted to exist by the quantum Gilbert-Varshamov bound \cite{PreskillQEC,  Calderbank_1996} and explicit or randomized constructions exhibiting such distance scaling have been investigated in the literature,  see,  e.g.,  refs.~\cite{BaconCircuit, Pantaleev_2021,Pantaleev_2022, Breuckmann_2021} and references therein. 
We define the analogous \textit{goodness} property 
for families of GKP codes.

\begin{mydef}[Good GKP codes]
A GKP code family $\CL_{n}\subset \mathbb{R}^{2n}$ parametrized by lattice dimension $2n$ with asymptotically non-vanishing rate \begin{equation}\lim_{n\rightarrow \infty}\log\det\lr{\CL_n}/n>0\end{equation} and distance scaling \begin{equation}\Delta^2=\Omega\lr{n}\end{equation} is \textit{good}.
\end{mydef}

The existence of such a family of good GKP codes has been established by Harrington and Preskill in refs.~\cite{HarringtonPreskill,  Harrington_Thesis}.  Their proof,  based on the existence of good GKP codes obtained from re-scaling symplectically self-dual lattices with shortest vector length $\lambda_1\lr{\CL}= \Omega\lr{\sqrt{n}}$,  whose existence had been shown by Buser and Sarnak \cite{Sarnak1994},  however,  is non-constructive.  In the following, we will review this construction of GKP codes,  which we have called \textit{scaled GKP codes} in 
ref.~\cite{Conrad_2022},  list some notable examples and show how the NTRU scheme \cite{Hoffstein_Pipher_Silverman_1998} yields a randomized construction of good GKP codes.

\section{Constructions of GKP codes}\label{sec:codes}

\paragraph{Scaled GKP codes} Central to the construction of GKP codes are the class of \textit{scaled GKP codes},  first introduced and analysed in refs.~\cite{GKP,  Harrington_Thesis},  where a GKP code is obtained by scaling a symplectically self-dual lattice (which we will refer to as \textit{symplectic lattice}) by a factor $\sqrt{\lambda}, \, \lambda \in \mathbb{N}$.  Let $\CL_0=\CL\lr{M_0}$ be such a $2n-$dimensional lattice,  where we choose $M_0$ as the symplectic basis \cite{lang2005algebra},  i.e.,  $M_0$ is such that
\begin{equation}
A_0=M_0JM_0^T=J.
\end{equation}
Clearly,  $A_0$ is integer.  Hence the scaled lattice $M=\sqrt{\lambda}M_0, \, \lambda \in \mathbb{N}$, 
also retains a symplectically integral Gram matrix
\begin{equation}
A=MJM^T=\lambda J.
\end{equation}
Such lattices are sometimes also called \textit{q-symplectic} \cite{Gama_2006} with $q=\lambda$.  The total encoded dimension and distance are
\begin{align}
D 			&=\sqrt{ |\det\lr{A}| }=\lambda ^n,  \\
\Delta 	&= \lambda ^{-\frac{1}{2}} \lambda _1\lr{\CL_0},
\end{align}
where $\lambda_1\lr{\CL_0}$ is the length of the shortest vector in the symplectic lattice $\CL_0$. 
The sympletic dual lattice is generated by $M^{\perp}=\lambda^{-\frac{1}{2}}M_0$.  If $M_0$ is stated in the symplectic basis one can immediately read off pairs of vectors with 
\emph{symplectic inner product}
\begin{equation}
\lr{\bs{\xi}^{\perp}_{i}}^TJ\bs{\xi}^{\perp}_{i+n}=\frac{1}{\lambda};  \; i=1,\dots, n 
\end{equation}
and symplectic inner product $0$ with any other row in $M_0$.
The corresponding displacement operators anti-commute up to phase $\omega_{\lambda}=e^{i\frac{2\pi}{\lambda}}$ and commute with each displacement associated to every other row of $M^{\perp}$, such that they form the logical generalized Pauli group
\begin{equation}
\overline{X}_i=D\lr{\bs{\xi}^{\perp}_{i}},\,  \overline{Z}_i=D\lr{\bs{\xi}^{\perp}_{n+i}};\hspace{.3cm}  \overline{X}_i^{\lambda},  \overline{Z}_i^{\lambda} \in \CS.
\end{equation}

Some examples of symplectically self-dual lattices are well known in the literature and also have been re-derived by exhaustive numerical search in ref.~\cite{Harrington_Thesis},  which we list in fig. \ref{fig:table_lattices} along with other symplectically integral lattices that produce notable GKP codes.  The smallest GKP lattices in that table include
\begin{itemize}
\item \textbf{The $\mathbb{Z}^2$} lattice with basis
\begin{equation}
M_{\mathbb{Z}^2}=I_2.
\end{equation}
This is also the \textit{symplectic basis} \cite{lang2005algebra}  for $\mathbb{Z}^2$,  i.e.,  it is such that $M_{\mathbb{Z}^2}JM_{\mathbb{Z}^2}^T=J$.
We refer to the scaling of this lattice or its $N$-fold direct sum by the factor $\sqrt{\lambda}=\sqrt{2}$ as the \textit{square GKP code}
\begin{equation}
\CL_{N\square}=\sqrt{2}\mathbb{Z}^{2N}.  
\end{equation}
It has already been noticed in 
ref.~\cite{GKP} that all Clifford operation for this code can be performed by means of symplectic operations.  Furthermore, 
ref.~\cite{Baragiola_2019} noticed that performing stabilizer measurements and corrective shifts on the vaccum produces the logical $\ket{H+}$ magic state vector.  This is due to the fact that $J_2$ is a symplectic orthogonal automorphism (the logical Hadamard) of the lattice.  This is by far the most simple and most popular GKP code discussed in the literature,  which also is owed to the fact the lattice is orthogonal such that decoding via CVP becomes a simple one-dimensional rounding protocol \cite{babai}.

\item \textbf{The hexagonal $A_2$ lattice} with symplectic basis
\begin{equation}
M_{A_2}=\frac{1}{\sqrt[4]{12}} \begin{pmatrix} 2 & 0 \\ 1 & \sqrt{3} \end{pmatrix} .
\end{equation}
This lattice is one of the \textit{root lattices} in $2$ dimensions and is known to yield the densest sphere packing there,  capable of tightly packing spheres of radius $\lambda_1\lr{A_2} / 2=\frac{1}{\sqrt[4]{12}}\approx 0.537\hdots$.  The scaling to a qubit-GKP code $\CL=\sqrt{2}A_2$ has distance
\begin{equation}
\Delta_{\sqrt{2}A_2} =\frac{1}{\sqrt[4]{3}}\approx 0.76\hdots,
\end{equation}
which is the highest distance one can obtain for a single-mode GKP code encoding a single qubit as it is the densest lattice packing in two dimensions \cite{ConwaySloane}.  Interestingly,  code states of the hexagonal GKP code have also been rederived in a numerical search for the most robust encoding of a qubit into an oscillator under photon loss \cite{Noh_Capacity}.  
\end{itemize}

Other high dimensional GKP codes built by scaling symplectic self-dual root lattices become increasingly complex and interesting,  such as the GKP code built upon the dihedral root system $D_4$ examined in ref.~\cite{Royer_2022},   the \emph{Gosset lattice} obtained from the exceptional Lie algebra $E_8$ or the Leech lattice $\Lambda_{24}$,  and warrant much further investigation in the future. 

\begin{figure*}
\center
\resizebox{\textwidth}{!}{
\begin{tabular}{c|c|c| c|c | c|c}
$n$ 	&	$\dim\lr{\CL_0} \,\lr{\CL}$				& $\CL_0$ 									& $\lr{\lambda_1\lr{\CL}}^2$	 		& Symp. self-dual										& Eucl.~self-dual 		& Concatenated (trivial sublattice)  \\ \hline
$1$ 	&	$2$									& $\mathbb{Z}^2$						& $1$ 												&\checkmark														& \checkmark						& -- \\
$1$	&	$2$									& $A_2$										&	$\frac{2}{\sqrt{3}}$						&\checkmark														& \checkmark						&	-- \\
$2$ 	& $4$									& $D_4$				& $\sqrt{2}$												&\checkmark \cite{Harrington_Thesis}								& \checkmark			&  $\CL_{\rm triv}\sim \Z^4$ w/ repetition code \cite{Royer_2022}\\ 
$4$	& $8$									& $E_8$										&$2$												&\checkmark														&\checkmark						& $\CL_{\rm triv}\sim 2\Z^8$ w/ Hamming code \cite{ConwaySloane} $\mathcal{H}_8 =[8,4,4]$\\
$6$	& $12$									& $K_{12}$				&$\frac{4}{\sqrt{3}}$	\cite{Harrington_Thesis}					&\checkmark \cite{Harrington_Thesis}	&\checkmark					& $ \CL_{\rm triv}\sim A_2^6$ \cite{Conway_Voroni}	\\
$12$&$24$									&$\Lambda_{24}$	& 4 \cite{ConwaySloane}			&\checkmark \cite{Sarnak1994}				&\checkmark					&$\CL_{\rm triv}\sim 2\Z^8$ w/ Golay code$^*$ \cite{ConwaySloane} $\mathcal{C}_{24}=[24,12,8]$ \\ \hline
$n$ & $2n$ & $\sqrt{\lambda/ q}L_{\rm NTRU}$ & $\Delta \sim \Omega\lr{\sqrt{n/\lambda}} $ & \checkmark  & \checkmark & $\CL_{\rm triv} \sim \sqrt{\lambda q}\Z^{2n}$ \\ \hline
$N$	&$2N$									&$\Lambda_{\square}\lr{\mathcal{Q}}$	& $\Delta\geq  \sqrt{d/2}$ & x												&x	& $\mathcal{Q}=[\![N,k, d]\!]$\\
$N$	&$2N$									&$\Lambda_{\mhexagon}\lr{\mathcal{Q}}$	& $\Delta=\sqrt{d/\sqrt{3}}$ & x				&x	&  $\mathcal{Q}=[\![N,k, d]\!]$

\end{tabular}}
\caption{Some notable weakly symplectically self-dual (symplectic) lattices that yield GKP codes.  The lower block indicates the concatenation of single mode $\CL_{\square}=\sqrt{2}\mathbb{Z}^2$ square GKP and $\CL_{\mhexagon}=\sqrt{2}A_2$ hexagonal GKP codes with qubit quantum error correcting- or detecting codes.  Note that concatenation with $\CL_{\mhexagon}$ does not formally produce a Construction A lattice,  but is related by a symplectic transformation $S_{\mhexagon}^n=\oplus_i^n S_{\mhexagon}$,  $S_{\mhexagon}=M_{A_2}^T$ to the concatenation with the square GKP code generated by $M_{\mathbb{Z}^2}=I_2$,  which in fact is Construction A. The symplectically self-dual root lattices listed in this table and their use as GKP codes have previously been identified in
ref.~\cite{Harrington_Thesis}. The re-scaled $L_{\rm NTRU}$ lattices that we use here to to construct NTRU-GKP codes are indicated between those and the ``more genuine" lattices corresponding to concatenated codes. The statements about (symplectic) self-duality are generally up to scaling and rotations.} \label{fig:table_lattices}
\end{figure*}


\paragraph{Concatenated GKP codes.} 
Once a logical qubit is obtained by means of a scaled GKP code,  one can concatenate it with a \emph{quantum error correcting}- or \emph{quantum error detecting code} (QECC/QEDC) $\mathcal{Q}=[\![N,k, d]\!]$ to obtain a GKP lattice with larger minimum distance.  If $Q$ is the set of symplectic vectors representing the stabilizers of a QECC/QEDC or generally,  any set of (binary) vectors such that $\forall p,q \in Q:\; p^T J q =0\mod\, 2$,  we can construct a  GKP lattice by means of \textit{Construction A} 
\cite{ConwaySloane,  Conrad_2022},  which yields
\begin{equation}
 \Lambda\lr{Q}:=\lrc{\bs{x}\in \mathbb{R}^{2n}\big\vert \sqrt{2}\bs{x}\mod 2 \in Q }.
\end{equation}
This lattice can be interpreted as the full-rank embedding of $Q$ into $\mathbb{R}^{2n}$ and inherits its main properties immediately from the code properties of $Q$ including decoding algorithms.  In refs.~\cite{Yang_thesis,  fukui2017analog,  toricGKP,  surfGKP,  Ravendraan2021,  Lin_2023},  known decoding algorithms for quantum error correcting codes such as \textit{minimum-weight-perfect-matching} (MWPM, which is an MED decoder) have been adapted to decode diverse concatenations of the single mode square GKP code with the popular \emph{surface},  \emph{toric},  \emph{color},  and \emph{quantum low-density-parity-check} (QLDPC) codes,  where the corresponding GKP-lattices can all be understood as Construction A lattices.  
As noted previously,  we denote the underlying multi-mode square-GKP code as $\CL_{N\square}$ and the full lattice corresponding to the concatenated code as $\CL= \Lambda\lr{Q}$,  such that we have $\CL_{N\square}\subseteq \CL$ and reversely $\CL^{\perp} \subseteq \CL_{N\square}^{\perp}$.
These procedures have in common that one decodes in two steps provided the syndrome and a generic correction $\bs{\eta}$: 1.  Solve CVP on the superlattice $\CL_{N\square}^{\perp}$,  perform the corresponding correction.  This step returns one to $ \CL_{N\square}^{\perp}$ and the residual syndrome is a genuine binary syndrome for $Q$.  2.  The qubit-level decoder solves some version of (approximate) CVP on $\CL^{\perp}$ provided that one starts out from $\CL_{N\square}^{\perp}$.  Before applying this decoder one computes a metric from the syndrome on  $\CL_{N\square}$ to take advantage of the continuous information held by the full GKP syndrome.  Finally one applies the qubit-level decoder with the amended metric.  In total,  this decoder can be pictured as a sequence
\begin{equation} 
 \mathbb{R}^{2n} \rightarrow  \CL_{N\square}^{\perp}  \xrightarrow{\mathtt{CVP}\lr{ \mu}} \CL^{\perp}.\label{eq:dec_conc}
\end{equation}

It is also known that one can solve \texttt{CVP} exactly on any Construction A lattice provided a soft decoder for the underlying binary code $Q$,  see ref.~\cite[p.~450]{ConwaySloane}.  This observation has, e.g.,  
been used to construct \texttt{CVP} algorithms for the $E_8$,  which can also be understood as a 
Construction A lattice on the $\mathcal{H}_8=[8,4,4]$ Hamming code.

\subsection{Decoding complexity of GKP codes}
Due to the lattice theoretic nature of GKP codes and their respective decoding problems and the well developed literature on lattice problems,  it is interesting to investigate the computational complexity of decoding GKP codes from this perspective.  Before we continue to construct GKP codes from a cryptosystem proposed for post-quantum cryptography in the next section,  we show that 1.  for GKP codes,  MLD decoding is at least as hard as MED decoding and 2.  MED decoding a concatenated (qubit-) GKP code implies a decoder for the corresponding qubit-code. 

We include these statements here because we find the proofs illustrative and wish to highlight that decoding complexity of GKP codes is an interesting question deserving of our attention.  Denote by \texttt{eMLD} the problem of evaluating the \texttt{MLD} probabilty given by the theta function on the RHS in eq.~\eqref{eq:MLD}.

\begin{lem}{($\mathtt{eMLD}\geq \mathtt{MED}$)}
    Given an oracle that evaluates $$\mathtt{eMLD}\lr{\bs{x}, \bs{\xi}^{\perp}, \CL, \overline{\sigma}}=\Theta_{\CL+\bs{\xi}^{\perp}+\bs{x}}\lr{\frac{i}{2\pi \overline{\sigma}^2}},$$ $\mathtt{CVP}\lr{\bs{x}, \CL^{\perp}}$ can be solved efficiently.
\end{lem}
\proof
Denote by $\mathtt{DecCVP}\lr{\bs{x}, \CL, r}$ the decisional CVP problem that outputs $\mathtt{True}$ if $\text{dist}\lr{\bs{x},\CL}\leq r.$  This is polynomially equivalent to the optimization- and search variants of $\mathtt{CVP}$ \cite{Regev_lecture}. 
First notice that we generally have
\begin{align} \Theta_{\CL^{\perp}+\bs{x}}\lr{\frac{i}{2\pi \overline{\sigma}^2}}
&=\sum_{\bs{\xi}^{\perp} \in \CL^{\perp}/\CL} \Theta_{\CL+\bs{\xi}^{\perp}+\bs{x}}\lr{\frac{i}{2\pi \overline{\sigma}^2}}
\nonumber
\\
&\geq e^{-\frac{1}{2\overline{\sigma}^2} \text{dist}\lr{\bs{x}, \CL^{\perp}}^2}.
\end{align} 
If $\mathtt{DecCVP}\lr{\bs{x}, \CL, r}$ is true,  then we further have 
\begin{equation}
e^{-\frac{1}{2\overline{\sigma}^2} \text{dist}\lr{\bs{x}, \CL^{\perp}}^2} \geq e^{-\frac{r^2}{2\overline{\sigma}^2} }
\end{equation}
for all $\overline{\sigma} \in \R$,  and hence we can solve $\mathtt{DecCVP}\lr{\bs{x}, \CL^{\perp}, r}$ by checking if above condition is true for sufficiently small $\overline{\sigma} < r$.
Alternatively,  w.l.o.g.  assume that $\CL \subset \Z^n$ and $\bs{x} \in \Z$.  Given access to $$\Theta_{\CL^{\perp}+\bs{x}}\lr{z}=\sum_{m\in \mathbb{N}} a_m e^{i\pi z m},$$ we can compute 
\begin{equation}
2 a_m=e^{m \pi \tau  }\int_{-1}^1 dt\, e^{-it\pi m}\Theta_{\CL^{\perp}+\bs{x}}\lr{t + i\tau}
\end{equation}
to evaluate $\{a_m\}$ for 
$m=1, \dots, M$,  where $M$ can be bounded by Mikowskis convex body theorem,  to find the smallest non-zero coefficient $a_m$.  This solves optimization-\texttt{CVP} which is polynomially equivalent to 
its search version.\endproof
Note that here we did not show that the full $\mathtt{MLD}$ problem
\begin{equation}
\mathtt{MLD}\lr{\bs{x}, \CL, \overline{\sigma}}=\argmax_{\bs{\xi}^{\perp} \in \CL^{\perp} / \CL}\Theta_{\CL+\bs{\xi}^{\perp}+\bs{x}}\lr{\frac{i}{2\pi \overline{\sigma}^2}}    
\end{equation}
is hard. 

In ref.~\cite[p. 450]{ConwaySloane},  it has (constructively) been shown that given a soft decoder for a binary code $C$, we can always solve $\mathtt{CVP}$ on the corresponding Construction A lattice $\Lambda\lr{C}$.  We explain this point, which also clarifies the geometric picture on decoding concatenated codes presented in eq.~\eqref{eq:dec_conc}.

\begin{lem}[\cite{ConwaySloane},
p.~450]
\begin{equation}
\mathtt{CVP}\lr{\cdot,\,\Lambda \lr{C}}=\mathtt{Decode\lr{C}}.
\end{equation}
\end{lem}

\proof
$C$ is embedded in $\Z^n$ by identifying the (scaled and shifted) Construction A lattice $\Lambda\lr{C}=1-2C + 4 \Z^n $,  where every bit string $\bs{b}\in C$ is mapped to $1-2\bs{b} \in \{-1,1\}^n$.  In this representation we consecutively solve $\mathtt{CVP}\lr{\cdot, 4\Z^n}$ and then apply the soft decoder for $C$,  which finds the closest transformed code word $\bs{c} \in 1-2C \in \{-1,1\}^n$ to input $\bs{x}' \in \mathbb{R}^n$.  As both decoders are exact,  with a little care (see ref.~\cite[p.~450]{ConwaySloane}),  this solves CVP exactly.  Note that the reverse direction is trivially true via the embedding of $C$ into $\mathbb{R}^n$ provided by Construction A and taking modulo $4\Z^n$.  A hard decoder,  that solves
\begin{equation}
\argmin_{\bs{c}\in C} d_H\lr{\bs{c}_b, \bs{x}_b} \label{eq:dec_classical}
\end{equation}
on binary input $\bs{x}\in \{-1,1\}^n$  is also derived from a soft decoder by noticing that $\|\bs{c}-\bs{x}\|_2^2=4 d_H\lr{\bs{c}_b, \bs{x}_b}$, where $\bs{x}_b$ represents the binary $\{0,1\}$ representation of $\bs{x}$ and $d_H$ is the Hamming distance.
\endproof

A strategy very similar to the one laid out in this proof has been e.g.  employed in ref.~\cite{Lin_2023} to decode the surface-GKP codes employing the minimum-weight-perfect-matching (MWPM) algorithm as soft decoder.

It is known that decoding classical error correcting codes, as specified in eq.~\eqref{eq:dec_classical} is a computationally hard problem in general \cite{Berlekamp, Vardy}.  Hardness of decoding qubit-based quantum error correcting codes has previously been investigated by reduction to the related problem on classical codes \cite{Hsieh_2011} to show its NP-completeness,  and by showing its relationship to computing weight enumerators of a linear code,  ref.~\cite{Iyer} has even shown its $\#P$-hardness in worst case complexity. 

By a similar line of argumentation, one notes that the coefficients of the shifted lattice theta function
\begin{equation}
 \Theta_{\CL+\bs{x}}\lr{z} = \sum_{m \in \Z} N_{m}\lr{\CL, \bs{x}} q^{m} \label{MLD_3},
\end{equation}%
with
\begin{equation}
    N_{m}\lr{\CL, \bs{x}}=\#\lrc{\bs{y} \in \CL+\bs{x}:\, \|\bs{y}\|^2=m }
\end{equation}
counting the number of lattice vectors such that $\dist\lr{\bs{x}, \CL}^2=m$ are hard to compute in general. 
While we do not attempt to complete such a proof here,  we expect a theta-function based analysis of the decoding complexity of quantum error correcting codes using concatenation with single-mode GKP codes to yield similar results as refs.~\cite{Hsieh_2011, Iyer}.

\section{GKP codes from NTRU lattices}\label{sec:NTRUGKP}

Random lattices and variations of lattice problems SVP and CVP play a prominent role in classical- and post-quantum cryptography due to their assumed hardness even for quantum computers in the worst-case,  as well as due to the feature of worst-case to average-case reductions for such problems \cite{Ajtai96}.  The proof of existence of what we termed \textit{good} GKP codes provided by 
ref.~\cite{HarringtonPreskill} can in essence be formulated using a Haar average over the moduli space of all symplectic lattices \cite{Sarnak1994}.  The analogous heuristic to lower bound the shortest vector in a general lattice is given by the Gaussian heuristic. 

\begin{GaussianHeuristic}
Let $L \subset \mathbb{R}^n$ be a sufficiently random full rank lattice with large $n$,  then we expect the smallest non-zero vector in the lattice will satisfy
\begin{equation}
\lambda_1\lr{L}\approx \sqrt{\frac{n }{2\pi e}} \det\lr{L}^{\frac{1}{n}}.
\end{equation}
\end{GaussianHeuristic}

\textit{Argument} \cite{Silverman_lecture, RandomLat}:
The moduli space of full rank lattices in $\R^n$ with unit covolume is given by $\CL_n=\mathrm{SL}_{n}\lr{\Z}\setminus \mathrm{SL}_{n}\lr{\R}$,  where the left\footnote{We write the left quotient because of the row-convention used in the definition of lattice bases.  In the literature one more commonly uses a right-quotient associated to a column-convention.} quotient $\mathrm{SL}_{2n}\lr{\Z}$ indicates the equivalence up to changes of basis.  There is a Haar measure $\mu_n$ over $\CL_n$,  normalized to $\mu_n\lr{\CL_n}=1$,  such that for Lebesque-integrable functions $f:\, \R^n \mapsto R$,  we have that \cite{macbeath_rogers_1958}

\begin{align}
    &\int_{L \in \CL_n} f\lr{L\setminus \lrc{0}}\, d\mu_n &= \int_{\R^n} f\lr{\bs{x}}\, d\bs{x}, \label{eq:lattice_design}\\
    &\, \text{where}\hspace{1cm } f\lr{L\setminus \lrc{0}} &= \sum_{\bs{x} \in L \setminus \lrc{0}} f\lr{\bs{x}}.\label{eq:ModuliAV}
\end{align}

Let $f\lr{\bs{x}}=\theta\lr{\|\bs{x}\|\leq R}$, where $\theta$ is the Heaviside function. Equation \eqref{eq:ModuliAV} yields
\begin{equation}
    \Big\langle \#\lrc{\bs{x}\in L:\, \|\bs{x}\|\leq R,\; \bs{x}\neq 0 }  \Big\rangle_{L\in \CL_n}=V_n(R),
\end{equation}
where 
\begin{equation}
V_n(R)=\frac{\pi^{\frac{n}{2}}}{\Gamma\lr{\frac{n}{2}+1}}R^n
\end{equation} is the volume of the centered $n-$ball $B_n(R)\subset \mathbb{R}^n$. 

We hence have that if lattices $L$ are sampled from a random distribution close to $ \mu_n$ in the moduli space of all lattices with $\det\lr{L} =1$,  the average number of non-zero lattice points of length at most $R$ is given by the volume of the $n$-ball,  $V_n(R)$.  Similarly,  it is reasonable to expect that the average number of non-zero lattice points of length at most $R$,  when the lattice has $\det\lr{L} \neq 1 $ and is sampled from an approximation to the Haar measure is given by $V_n\lr{R}/\det\lr{{L}}$.  

Using Stirling's approximation, the smallest $R$ for which this number becomes non-zero is given by $R\approx \sqrt{n/2\pi e}\det\lr{L}^{\frac{1}{n}}$.
\qed

We remark that the Gaussian Heuristic is a statement accepted to be generally true in lattice theory and post-quantum cryptography.  In the above ``proof sketch" the ``heuristic" enters in the assumption that the design property eq.\  \eqref{eq:lattice_design} still holds for measures that only approximate the Haar measure on the space of lattices and that it moreover also still holds when the lattices are not of $\det\lr{\CL}=1$. 

The Gaussian heuristic motivates that lattices with $\lambda_1=\Omega\lr{\sqrt{n}}$ can be found amongst sufficiently random sets of lattices.  Buser and Sarnak \cite{Sarnak1994} showed that there is also a Haar measure over the moduli space of \emph{symplectic} lattices,  using which Harrington and Preskill identified the existence of good GKP codes by a similar calculation as presented above \cite{HarringtonPreskill}. 
\\

The construction of random lattices is a crucial ingredient to lattice based cryptography.

In this section we introduce the NTRU cryptosystem and show that random \emph{NTRU lattices} obtained from variations of the NTRU cryptosystem are in fact symplectic,  such that they allow to construct GKP codes as scaled GKP codes.  We discuss scenarios where NTRU lattices are sufficiently random to follow the Gaussian heuristic or,  at least,  can be shown to admit a lower bound $\lambda_1 \geq \Omega(\sqrt{n})$ with high probability.

The so-derived GKP codes share characteristics of both scaled- and concatenated GKP codes.  These NTRU lattices have been originally formulated in the cryptanalysis of attacks on the NTRU cryptosystem \cite{Hoffstein_Pipher_Silverman_1998,  Coppersmith_shamir,  May1999,  MayCryptanalysis} and their symplecticity has been motivation to further the study of lattice reduction algorithms for symplectic lattices \cite{Gama_2006}.  As GKP codes,  these lattices are particularly interesting as they can be understood as certain generalization of \textit{cyclic} quantum error correcting codes such as the well known $XZZX-\,  [\![5,1,3]\!]$  quantum error correcting code \cite{Bennett_1996} or the repetition code and have a similar algebraic basis as the recently introduced lifted product codes \cite{Pantaleev_2021}.

\subsection{The NTRU cryptosystem}

We describe the NTRU cryptosystem to the degree necessary to understand the structure of the corresponding lattices and GKP codes constructed here.  For more detail we refer the reader to the cited literature. The presentation here is largely derived from the presentations in refs.~\cite{Hoffstein_Pipher_Silverman_1998,  May1999,  MayCryptanalysis, Bernstein_Buchmann_Dahmen_2009, StehleSteinfeld, Halevi_lecture}. 

The NTRU cryptosystem is most natively formulated using polynomial rings $R=\Z\lrq{x} /\Phi$,  where we will take the quotient $\Phi=x^n+\phi_{n-1}x^{n-1}+\hdots+\phi_0$ as $\Phi=\Phi_0:=x^n-1$ in the bulk of this paper,  as used in the original description of the (heuristically secure) NTRU cryptosystem \cite{Hoffstein_Pipher_Silverman_1998}.  We will keep $\Phi$ general whenever possible to be able to discuss the provably secure version of the NTRU cryptosystem \cite{StehleSteinfeld} with irreducible $\Phi=x^n+1$ later.  We denote $R_q=R/qR$ with a typically large modulus parameter $q$ and $R_p=R/pR$ with a typically small $p$ coprime with $q$.  Whenever we take the modulus, $\mod q$ or $\mod p$, we refer to the (coefficient-wise) reduction into the centered fundamental domains $\lrq{-\frac{q}{2}, \frac{q}{2}}$ resp.~$\lrq{-\frac{p}{2}, \frac{p}{2}}$.

We denote multiplication in $R$ as $fg \mod \Phi,\, f,g\in R $, where we assume the reduction $\mod \Phi$ ($\mod q/p$) to be implicit by specifying the image and use a bold $\bs{f}=\mathrm{coeff}(f)$ to refer to the  coefficient vector $\bs{f}=(f_0, f_1, \dots, f_{n-1})$ of $f\in R$ (note that any polynomial in $R$ can be represented with $n-1$ coefficients, for that every power $x^n$ can be replaced by $x^n-\Phi$ when working over $\mod \Phi$. 

Denote the uniform distribution of polynomials $p\in R_q$ with $d_1$ coefficients $+1$, $d_2$ coefficients $-1$ and $n-(d_1+d_2)$ coefficients $0$ as $D\lr{d_1, d_2}$. Further denote the set of invertible elements in $R_q$, 
i.e.,  elements $f$ for which $f^{-1}\in R_q$ exists,  as $R_q^{\times}$.

The NTRU cryptosystem, specified by parameters $\lr{n, \Phi, d, q, p }$ operates as follows:

\begin{enumerate}
    \item \textbf{Key generation}: Sample $\tilde{f}\hookleftarrow  D(d,d)$ until $f=1+p\tilde{f}\in R_q^{\times}$,  sample $\tilde{g} \hookleftarrow  D(d,d)$ to obtain $g=p\tilde{g}$.  Return the secret key pair $(f,g)$,  and the public key $h=gf^{-1} \in R_q$.
    \item \textbf{Encryption}: Given the public key $h \in R_q$ and a message $m \in R_p$,  sample a random polynomial $r\in R_p$ and compute the ciphertext $c=hr+m \in R_q$.
    \item \textbf{Decryption}: Given the ciphertext $c$ and secret key $f$,  compute $cf \mod q \mod p= gr+fm \mod q \mod p = m \in R_p$.
\end{enumerate}

The secret key polynomials $(f,g) \in R_q^{\times}\times R_q$ are by construction such that $f \mod p=1$ and $g \mod p=0$.  Decryption is guaranteed to be successful whenever all the coefficients involved are sufficiently small,  such that $cf  = gr+fm $ holds as equality in $R$, and not just merely in $R_q$ \cite{Silverman_lecture}.

\subsection{Symplectic ideal and NTRU lattices}

The security assumption underlying this cryptosystem as the in-retrievability of the secret key is the hardness of the polynomial factorization problem in $R_q$ and secret key retrieval attacks have been formulated already in early analyses of the NTRU cryptosystem \cite{Hoffstein_Pipher_Silverman_1998, Coppersmith_shamir, MayCryptanalysis}.

\begin{Assumption}[Polynomial factorization problem \cite{MayCryptanalysis}]Given a polynomial $h=f^{-1}g \in R_q$ where the coefficients are small compared to q.  For suitable parameter settings it is intractable to find small polynomials $f' ,g'\in R_q$ such that $f'  h=g'\in R_q$.\label{ass:PFP}
\end{Assumption}

 Under the premise that the coefficient vectors of the secret key $(f,g)$ are \textit{short},  a typical attack is formulated as the task of finding short polynomials $(f', g') \in R_q^2$ such that $fh=g \in R_q$, where the length of the polynomial pair is defined as the norm of their joined coefficient vectors $\|(f', g')\|_l=\|(\bs{f'}^T , \bs{g'}^T )\|_l$.  We will use the $l=2$ norm unless specified otherwise. 
 The attack is carried out by defining the NTRU lattice as an $R$-module $L_R\subseteq R^2$,  which admits a basis in its Hermite normal form
\begin{equation}
    H_R=\begin{pmatrix}
        1 & h \\ 0 & q
    \end{pmatrix}.\label{eq:H_R}
\end{equation}
Elements of the $R$-lattice are of the form 
\begin{align}
    (f'\; u)H_R&=(f'\; u)\begin{pmatrix}
        1 & h \\ 0 & q
    \end{pmatrix} \\
    &=(f'\; f'h+uq ) \nonumber\\
    &=(f', f'h+uq) ,\; (f',u) \in R^2,\nonumber 
\end{align}
each of which represent admissible solutions to the equation $f'h=g' \in R_q$,  such that short vectors in $L_R$ are expected to correspond to the NTRU secret key pair.  In a more general classification, one can view the $R$-lattice $L_R=L_R(h)$ as an rank-$2$ ideal lattice \cite{compactknapsacks},  corresponding to the principal ideal $I=\langle h\rangle \subseteq R$.

$H_R$ is, in fact, also a $q$-symplectic matrix in $R^{2\times 2}$,  with respect to the symplectic form
\begin{equation}
J_R=    \begin{pmatrix}
        0 & 1 \\ -1 & 0
    \end{pmatrix}\in R^{2\times 2},
\end{equation}
with
\begin{equation}
    H_RJ_RH_R^T=\begin{pmatrix}
        h^T-h & q \\ -q & 0
    \end{pmatrix}
    =qJ_R \in R^{2\times 2}
\end{equation}
because $h$ is a scalar in $R$. 

Analyses of lattices and associated algorithms are typically formulated over $\Z$-lattices,  where linear combinations of basis vectors are taken with integer coefficients.  We map the rank-$2$ $R$-lattice $L_R$ to a rank-$2n$ $\Z$-lattice $L\subseteq \Z^{2n\times 2n}$ by defining a homomorphism onto a $\Z^{n \times n}$ matrix
\begin{align}
    C_{\Phi}:\; 
    R&\rightarrow \Z^{n \times n} \\
    f &\mapsto C_{\Phi}(f), \\ 
    \lr{C_{\phi}\lr{f}}_{i,j}&=(T_{\Phi}^i\bs{f})_j,\, i,j=0,\dots ,n-1,
\end{align}
where the rows are given by the vectors $T_{\Phi}\bs{f}$ and
    \begin{equation}
        T_{\Phi}=
        \begin{pmatrix}
            \bs{0}^T & -\phi_0 \\
            I_{n-1} & -\bs{\phi}_{1:n-1}
        \end{pmatrix}
    \end{equation}
implements the map $f\mapsto xf \mod \Phi \in R$ on the coefficient vector $\bs{f}$ of $f$ by left multiplication.

$C_{\Phi}$ is linear over $\Z$,  such that we can express the homomorphism on every polynomial $f\in R$ as
\begin{align}
    C_{\Phi}\lr{f}
    &=\sum_{i=0}^{n-1}f_i C_{\Phi}\lr{x^i}\\
    &=\sum_{i=0}^{n-1}f_i C_{\Phi}\lr{1}\lr{T_{\Phi}^T}^i,
    \nonumber 
    \\
    &=\sum_{i=0}^{n-1}f_i\lr{T_{\Phi}^T}^i,\nonumber
\end{align}
where we have used that $C_{\Phi}\lr{1}=I_n\; \forall \Phi$.
In this representation,  it is evident that $C_{\Phi}\lr{f}$ acts via right action
\begin{equation}
    \mathrm{coeff}(fg \mod \Phi)=\bs{f}^TC_{\Phi}\lr{g}=\bs{g}^TC_{\Phi}\lr{f}
\end{equation}
and that 
\begin{equation}
    C_{\Phi}\lr{f g \mod \Phi}= C_{\Phi}\lr{f} C_{\Phi}\lr{g}
\end{equation}
indeed represents a homomorphism.  When $\Phi=\Phi_0=x^n-1$,  $C_{\Phi}\lr{f}$ is simply the (row) circulant matrix of the coefficient vector $\bs{f}$.  Circulant matrices are not symmetric,  but have a mirror symmetry along the anti-diagonal,  $R_n C_{\Phi_0}^T\lr{f} R_n=C_{\Phi_0}\lr{f},$ where $R_n$ is the anti-diagonal matrix $(R_n)_{i,j}=\delta_{i, n-1-j}, \, i,j=0,\dots, n-1$.
We also define a related map
\begin{align}
    A_{\Phi}:\; 
    R&\rightarrow \Z^{n \times n}, \\
    f &\mapsto A_{\Phi}(f), \\ 
    \lr{A_{\phi}\lr{f}}_{i,j}&=(T_{\Phi}^{-i}\bs{f})_j,\, i,j=0,\dots ,n-1,
\end{align}
where $T_{\Phi}^{-i}=\lr{T_{\Phi}^{-1}}^i$ and for $\Phi=\Phi_0=x^n-1$ this is the symmetric anti-circulant matrix of the coefficient vector $\bs{f}$,  $A_{\Phi_0}^T(f)=A_{\Phi_0}(f)$.  Since $A_{\Phi}$ is also $\Z$-linear, here we have
\begin{align}
    A_{\Phi}\lr{f}
    &=\sum_{i=0}^{n-1}f_i A_{\Phi}\lr{x^i}\\
    &=A_{\Phi}\lr{1}C_{\Phi}\lr{f},\nonumber
\end{align}
where, for $\Phi=\Phi_0=x^n-1$, 
we have that 
    \begin{equation}
        A_{\Phi}\lr{1}=
        \begin{pmatrix}
            1 & 0 \\
            \bs{0} & \overline{I}_{n-1}
        \end{pmatrix}=:\sigma
    \end{equation}
is the orthogonal coefficient mirror $\sigma=\sigma^T$ that maps the coefficient vector $f(x) \in R$ to that of $f\lr{x^{-1}}=f\lr{x^{n-1}}\in R$ \cite{MayCryptanalysis}.

The so-defined maps allow us to map the earlier defined $R$-lattice $L_R$ onto a lattice $L=L(h)\subseteq \Z^{2n}$ by applying the corresponding homomorphism on the entries of the basis

\begin{equation}
    H_R=\begin{pmatrix}
        1 & h \\ 0 & q
    \end{pmatrix}
    \mapsto 
    H=\begin{pmatrix}
        I_n & C_{\Phi}\lr{h} \\ 0 & q I_n
    \end{pmatrix}.
\end{equation}
It can be checked that the lattice spanned by the basis contains all secret key pairs $(\bs{f'}^T\; \bs{g'}^T)$ corresponding to solutions $fh=g \in R_q$.  It is however not symplectic.  For $\Phi=\Phi_0$ we, however,  have that 
\begin{equation}
    H^{cs}=\begin{pmatrix}
        I_n & A_{\Phi_0}\lr{h} \\ 0 & q I_n
    \end{pmatrix}=\begin{pmatrix}
        I_n & \sigma C_{\Phi_0}\lr{h} \\ 0 & q I_n
    \end{pmatrix}\label{eq:HCS}
\end{equation}
is indeed symplectic and corresponds to a rotation of the lattice $L$,
\begin{equation}
    (\sigma \oplus I_n)H^{cs} = H_{\Z}(\sigma^T \oplus I_n),
\end{equation}
since $(\sigma \oplus I_n)$ is unimodular.  This is the basis used by Coppersmith and Shamir in their attack on the NTRU cryptosystem \cite{Coppersmith_shamir, MayCryptanalysis}.  We generalize this observation to the following statement.

\begin{lem}
    An NTRU lattice $L\subseteq \Z^{2n} \subset \R^{2n}$ given by generator
    \begin{equation}
         H_{\Z}=\begin{pmatrix}
        I_n & C_{\Phi}\lr{h} \\ 0 & q I_n
    \end{pmatrix}
    \end{equation}
    is equivalent to a $q$-symplectic lattice $L'\subset \R^{2n}$ for all $h$ if
    there exists a signed permutation matrix $\sigma_{\Phi} \in \Z^{n \times n} \cap O\lr{n}$ such that 
    \begin{equation}
        \lr{\sigma_{\Phi}C_{\Phi}\lr{h}}^T=\sigma_{\Phi}C_{\Phi}\lr{h}
    \end{equation}
    is symmetric.
\end{lem}
\proof 
A lattice generated by $M$ is equivalent to a lattice generated by $M'$, such that $\det M=\det M'$ if and only if there exists a unimodular matrix $U$  and an orthogonal matrix $O$ such that $M'=UMO$ \cite{ConwaySloane}.  Take $O=\lr{\sigma_{\Phi}^T\oplus I_n}$ and $U=\lr{\sigma_{\Phi} \oplus I_n}$.
\endproof
\begin{cor}
    NTRU lattices over $\Phi=\Phi_0$ and $\Phi=x^n+1$ are equivalent to $q$-symplectic lattices
\end{cor}
\proof
For $\Phi=\Phi_0$ we already saw earlier that $\sigma_{\Phi_0}=\sigma$ provides a symmetric matrix $\sigma C_{\Phi}\lr{h}$ for all $h$.  For $\Phi=x^n+1$ this is also the case, with
   \begin{equation}
        \sigma_{\Phi}=
        \begin{pmatrix}
            1 & \bs{0}^T \\
            \bs{0} & -\overline{I}_{n-1}
        \end{pmatrix}=A_{\Phi}\lr{1}
    \end{equation}
and $A_{\Phi}\lr{1}C_{\Phi}\lr{h}=A_{\Phi}\lr{h}$ is such that the first row is $\bs{h}^T$ and every other row is generated by permuting the first element around the ``periodic boundary" on the right to the left while adding a $-1$ factor.  This matrix is clearly symmetric and $\sigma, \sigma_{\Phi}$ are signed permutations.
\endproof
Finally,  the fact that these NTRU lattices $L$ corresponds to ideals $I=\langle h\rangle \subseteq R$ equips them with the symmetry $L=(T_{\Phi}\oplus T_{\Phi})L$.  When $\Phi=\Phi_0$ we have that the symmetry is $n$-fold, $T_{\Phi_0}^n=I$ and similarly for $\Phi=x^n+1$ we have $T_{\Phi}^n=-I$. 

Henceforth we will default to $\Phi=\Phi_0$ unless otherwise specified and omit the corresponding $\Phi_0$ index from $C_{\Phi_0}$ and $A_{\Phi_0}$.  The anti-circulant matrix $A\lr{h}=\sigma C\lr{h}$ implements a homorphism from $R$ with respect to a modified matrix multiplication
\begin{equation}
    A\lr{f}\sigma A\lr{h}=\sigma C\lr{f}C\lr{f}=A\lr{fg}.
\end{equation}
We denote $A\lr{f}\sigma=:A^{\sigma}\lr{f}$,  such that $A^{\sigma}\lr{f}A\lr{g}=A\lr{f}$.

On $\Z^n$, ciphertexts produced by the NTRU encryption with secret key pair $(f, g)$ and public key $h$ take the form
\begin{align}
    \bs{c}^T
    &=\bs{m}^T+\bs{r}^T C(h)\,  \mod q\\
    &=\bs{m}^T+\lr{\sigma \bs{r}}^T  \sigma  C(h)\, \mod q
    \nonumber
\end{align}
and decryption is carried out by left-multiplying with $A^{\sigma}\lr{f}$ and reducing $\mod q$ and $\mod p$.

The corresponding $q$-symplectic generator of the underlying lattice is given by
\begin{equation}
    H=\begin{pmatrix}
        I_n & A\lr{h} \\ 0 & qI_n
    \end{pmatrix},\label{eq:HCS}
\end{equation}
which is already a $q$-symplectic basis for the weakly symplectically self-dual lattice $L$.

We use this lattice as starting point to define a scaled GKP-code by taking $\CL=\sqrt{({\lambda}/{q})}L$ with generator 
$M=\sqrt{({\lambda}/{q})}H$.  Similar to the discussion earlier,  the GKP code built this way will encode $D=\lambda^n$ logical dimensions with symplectic dual \begin{equation}\CL^{\perp}=L/\sqrt{\lambda q}\end{equation} and  distance \begin{equation}\Delta=\lambda_1\lr{L}/\sqrt{\lambda q}.\end{equation}
For randomly chosen $f,  g \in \CR$,  the Gaussian heuristic gives an estimate for the shortest vector length and has been used in the original NTRU work to argue about the security of the scheme \cite{Hoffstein_Pipher_Silverman_1998}. If the Gaussian heuristic were to hold, the so-constructed NTRU lattices would attain parameters

\begin{align}
D &= \lambda^n,\\
\Delta &\geq  \sqrt{\frac{n }{\lambda \pi e}},
\end{align}
which are \textit{good}.  

However,  the Gaussian heuristic does not always hold for NTRU lattices with arbitrary parameters.  Due to the the sub-lattice $q\Z^{2n}\subset L$ there always exist trivial vectors $q \bs{e}_i,\, i \in \lrq{1,2n}$ of length $q$ in  $ L$ which yield logically non-trival vectors of length $\sqrt{q/\lambda}$.  A shortest vector length $\lambda_1\lr{L}$ growing with $\sqrt{n}$ can however be maintained by choosing suitably large $q$ scaling with $n$.  Furthermore,  NTRU lattices (with $\Phi_0$) are constrained by 1.  being \textit{cyclic} lattices and 2.  having an existing inverse of $f \in R_q$ and 3.  having a fixed number $2d$ of non-zero coefficients in the vector corresponding to the secret key $(\sigma\lr{\bs{f}}^T,\bs{g}^T)^T \in L$,  which on the one hand make it not immediately clear if they would be sufficiently random for the Gaussian heuristic to hold,  and on the other hand already present short vectors of length $\leq  O(\sqrt{d})$.  These points have been addressed in refs.~\cite{Qi_earchive, Jingguo_Qi}, where the authors show the following statement.

\begin{cor}[{\cite[{Corollary 3}]{Qi_earchive, Jingguo_Qi}}]\label{cor:Qi}
If $d = \lfloor n/3 \rfloor $, then with probability greater than $1-2^{-0.1n}$ the shortest vector in a random NTRU lattice has length greater than
$\sqrt{0.28 n}$.
\end{cor}

This statement gives us confidence to claim that random NTRU lattice based GKP codes as constructed above can be expected to be \textit{good} when the parameters are chosen properly, as summarized by the following.
\begin{prop}[Good codes from NTRU lattices]
\label{prop_1}
    A GKP code with $\CL=\sqrt{({2}/{q})}L$, where $L$ is the NTRU lattice over $\Phi_0$ specified in the basis eq.~\eqref{eq:HCS} with $d = \lfloor n/3 \rfloor $ encodes
    \begin{equation}
        k=n
    \end{equation}
    qubits and has with probability greater than $1-2^{-0.1n}$ a distance given by
    \begin{equation}
        \Delta= \min \lrc{\sqrt{\frac{0.14 n}{q}}, \sqrt{\frac{q}{2}}}.
    \end{equation}
    For sufficiently large constant $q$ and $n\leq q^2/0.28$ this defines a randomized family of \textit{good} GKP codes.
\end{prop}
\proof
Follows immediately from corollary \ref{cor:Qi} and the GKP-code construction laid out in the main text.
\endproof

\subsection{Numerical results}
\begin{figure*}
 \center
 \includegraphics[width=\textwidth]{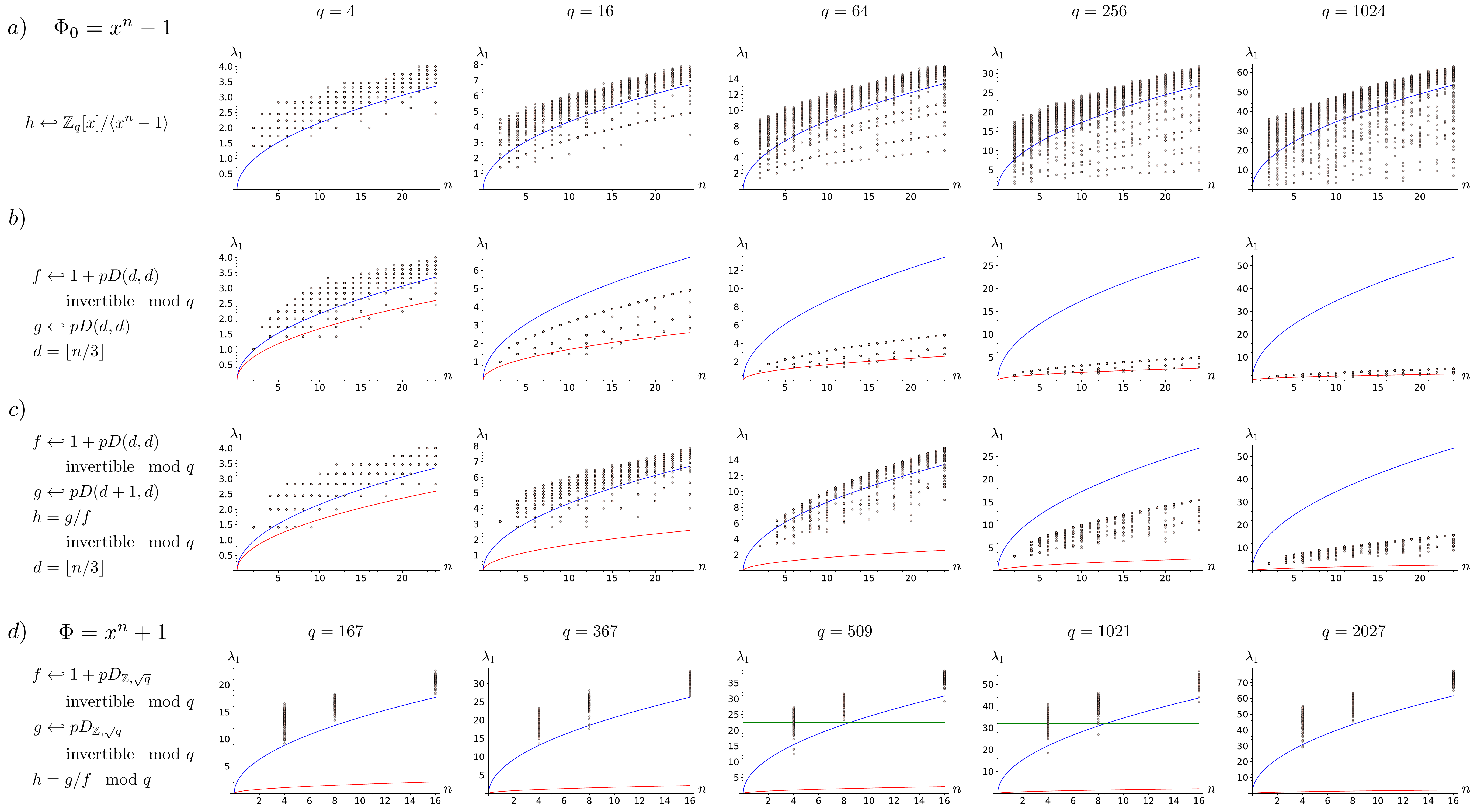}
 \caption{ Shortest vector lengths computed via full \texttt{HKZ} reduction of $a)$ random cyclic ($\Phi_0=x^n-1$) lattices as generated by the hard lattice generator in \texttt{sagemath}, $b)$ random NTRU lattices with $p=3$ and $d=\lfloor n/3 \rfloor$ and $c)$ random NTRU lattices where $h$ is invertible in $R_q$  for varying $q=2, \dots , 2048$.  In $d)$ we sample NTRU lattices generated with the irreducible quotient $\Phi=x^n+1$, where $n$ is a power of 2. 
 For each $n \in [2,24]$ we sample $100$ NTRU lattices and compute the shortest vector by computing the HKZ reduced lattice basis.  For reference, we plot the expected shortest vector length from the Gaussian heuristic $\lambda\lr{n}=\sqrt{nq/\pi e}$ in blue and the expected lower bound $\lambda_0\lr{n}=\sqrt{0.28 n}$ in red.  In panel $d)$,  we have also included a green line at $\sqrt{q}$,  which is the standard deviation of the discrete Gaussian distribution $f, g$ are sampled from and is related to a probabilistic lower bound for $n\geq 8$ a power of $2$ on the shortest infinity norm $\lambda_1^{\infty}\lr{L}$ derived in ref.~\cite{StehleSteinfeld}.
 The \texttt{sagemath} \cite{sagemath} code as well as all numerical data presented here is available under ref.~\cite{GitLink}. The \texttt{sagemath} functionalities to construct NTRU lattices are partially adapted from ref.~\cite{LatticeHacks}.} \label{fig:NTRU_sample}
 \end{figure*}

In fig.~\ref{fig:NTRU_sample},  we have computed the shortest vector lengths for $N_{\rm sample}=100$ randomly sampled NTRU lattices for varying $q$ and $n$ with $p=3$.  We compare samples over NTRU-like random cyclic lattices,  where $h$ is sampled randomly from $R_q$ in row $a)$ with NTRU lattices over $\Phi=x^n-1$ with $f$ invertible in $R_q$ and bounded non-zero entries $d=\lfloor n/3 \rfloor$ (in row $b)$).  We also compare the average length of shortest vectors for even more constrained NTRU lattices where we also required the public key $h$ to be invertible in $R_q$ in row $c$.  In this case we obtain $g$ from the amended distribution $g \sim p D(d+1,d)$ since otherwise $g$ -- and thus $h$ -- would have a trivial root $g(1)=0$ rendering the polynomial non-invertible.
Finally,  in row $d)$,  we perform the experiment using the setup of ref.~\cite{StehleSteinfeld},  where the quotient $\Phi=x^n+1$ is chosen to be irreducible and $n$ is a power of $2$. 

In our statistics we observe that random cyclic lattices (row $a$) appear to agree well with the Gaussian heuristic, while the growth of the shortest vector length of the NTRU lattices in row $b)$ and $c)$ degrades with increasing $q$,  consistent with the  bound given in Corollary \ref{cor:Qi}.  Based on our numerics,  we also conjecture the following.

\begin{conjecture}[Good GKP codes]\label{conj:random_h}
    A GKP code with $\CL=\sqrt{{\lambda}/{q}}L$, where $L$ is specified by the basis in \eqref{eq:HCS} and $h$ is selected at random from $R_q=\Z_q\lrq{x}/\langle x^n-1\rangle$,  is likely a good code with $k=n$ and 
    \begin{equation}
        \Delta \geq \min\lrc{ \sqrt{\frac{n }{\lambda \pi e}}, \sqrt{\frac{q}{\lambda}}}.
    \end{equation}
\end{conjecture}

Finally,  in row $d)$ we observe a good agreement of the shortest vector lengths with the scaling proposed by the Gaussian heuristic.  In ref.~\cite{StehleSteinfeld} a probabilistic lower bound on the smallest infinity norm $\lambda_1^{\infty}\lr{L}$ has been proven,  which we include in the figure.  As we will discuss later in the manuscript,  GKP codes derived from this particular NTRU setup is is of cryptographic relevance. Based on our numerical observations we hence also conjecture that such GKP codes are likely to be good.

\begin{conjecture}[Good GKP codes]\label{conj:SS}
    A GKP code with $\CL=\sqrt{{\lambda}/{q}}L$, where $L$ with $\det L = q^n$ is equivalent to  NTRU lattice specified by the basis in \eqref{eq:HCS} and $h=g/f\leftarrow f, g$ are sampled  at random from a Gaussian distribution with variance $\sigma^2=q$ in  $R_q=\Z_q\lrq{x}/\langle x^n+1\rangle$, $q\geq \mathtt{poly}(n)$ and $n\geq 8$ a power of $2$ is likely a good code with $k=n$ and 
    \begin{equation}
        \Delta \geq  \sqrt{\frac{n }{\lambda \pi e}}.
    \end{equation}
\end{conjecture}

In contrast to the previous statement in proposition \ref{prop_1}, these distance bounds do not suffer from choosing larger modulus $q$, but we can pick it arbitrarily large to obtain high distances. 

The trivial sub-lattice $L_q=q\Z^{2n}\subseteq L$ which enforces the $q$ modularity in the cryptographic setup is analogue to the structure of concatenated (hypercubic) GKP codes $\CL_{\rm triv}=\sqrt{\lambda q}\Z^{2n} \subseteq \CL$,  such that the lattices $\CL$ defined above may be interpreted as a concatenated (qudit) GKP code where $\CL_{\rm triv}$ defines the underlying single mode qudit-code with $D=\lambda q$.  It is interesting that this class of NTRU-GKP codes thus shares characteristics of both \textit{scaled-} as well as \textit{concatenated} GKP codes. 

\section{Decoding GKP codes from NTRU lattices}\label{sec:decodingNTRU}
A code state that (either through a natural error process or by deliberate modification) undergoes a displacement by
\begin{equation}
\bs{e}=\begin{pmatrix}
\bs{x} \\ \bs{y}
\end{pmatrix}
\end{equation}
gives rise to trivial syndrome

\begin{align}
\bs{s}_{\rm triv}&=\sqrt{\lambda q}\bs{e} \mod 1.
\end{align} Due to the simple orthogonal structure of $\CL_{\rm triv}$ a first step of the correction is easily carried out by applying the correction $\bs{\eta}=-\bs{s}_{\rm triv}/\sqrt{\lambda q} $.
After correcting for the trivial syndrome (associated to the underlying hypercubic GKP code) the remaining error is the unknown,  but likely short,  vector 
\begin{equation}
\bs{e}'=
\frac{1}{\sqrt{\lambda q}}
\begin{pmatrix}
\bs{u} \\ \bs{v} 
\end{pmatrix} \in \CL_{\rm triv}^{\perp},\; \bs{u},  \bs{v} \in \Z^n.
\end{equation}
The residual error can be considered as living on the scaled $q$-ary lattice 
\begin{equation}
\CL_q=\frac{1}{\sqrt{\lambda q}}\mathbb{Z}^{2n}_q
\end{equation}
dual to the trivial stabilizer lattice  and has a probability distribution induced by the trivial syndrome and correction 
\begin{equation}
P\lr{\bs{e}'}\propto \sum_{\bs{t} \in \CL_{\rm triv}} e^{-\frac{\lr{\bs{e}'+\bs{s}_{\rm triv}+\bs{t}}^2}{2\overline{\sigma}^2}}. \label{eq:res_prob}
\end{equation}
The remaining syndrome is
\begin{align}
\bs{s}&=MJ\bs{e}' \mod 1 \nonumber \\
&= \frac{1}{q}\begin{pmatrix}
\bs{v}-A\lr{h}\bs{u} \mod q \\ 0 \mod 1 
\end{pmatrix} .
\end{align}
In the first block of the syndrome $q\bs{s}_1=\bs{v}-A\lr{h}\bs{u} \mod q$ we recognize the structure of the NTRU ciphertext.  The position of the message is taken by $\bs{m}=\bs{v}$ and the random vector is replaced by $\bs{r}=-\sigma(\bs{u})$.  Following the standard NTRU decryption process now allows to obtain $\bs{v} \mod q \mod p$ as well as 
\begin{equation}\bs{u}=qA^{\sigma}\lr{h^{-1}}(\bs{v}-q\bs{s}_1) \mod q\end{equation} 
when $h$ is also chosen to by invertible in $R_q$ with inverse $h^{-1}$.  When $p$ is not prime,  we can instead obtain $\bs{u} \mod q \mod p_i$ for each prime factor $p_i$ of $p$ and estimate $\bs{u} \mod q $ via the Chinese remainder theorem.  We refer to this decoding routine as \texttt{NTRUDecode} and provide some small scale numerical tests in
appendix \ref{app:numerics}.

Alternatively,  we can decompose the remaining syndrome as

\begin{equation}
q\bs{s}=\begin{pmatrix}\bs{v} \\ -\bs{u} \end{pmatrix} + \underbrace{\begin{pmatrix}-A\lr{h}\bs{u} \\ \bs{u} \end{pmatrix}}_{\in  \CL_{\rm cs}^J}, \label{eq:decomp}
\end{equation}
where the vector on the RHS is element of the flipped NTRU lattice generated by the public basis
\begin{equation}
H^J=\begin{pmatrix}
q I & 0 \\ -A\lr{h} & I
\end{pmatrix}.
\end{equation}

Equation \eqref{eq:decomp} shows that a likely,  i.e.,  small,  error vector $\begin{pmatrix}\bs{v} \\ -\bs{u} \end{pmatrix}$ can indeed be obtained by solving $\mathrm{CVP}\lr{\CL^J,  q\bs{s}}$.  In appendix \ref{app:numerics},  we implement an approximation of this CVP instance using Babai's nearest plane algorithm with the HKZ reduced flipped public basis as \texttt{BabaiDecode} again only for small parameters $n$.

We recognize that the solving CVP or BDD on the respective NTRU lattices provides a viable route to decoding.  To analyze how well decoding can be carried out efficiently when provided only the public- vs.~the secret basis,  we analyze the radius $\epsilon$, up to which $\mathtt{BDD}_{\epsilon}$ can be implemented given the respective bases. 

\subsection{Bounding $\mathtt{BDD}_{\epsilon}$}

The maximum $\mathtt{BDD}_{\epsilon}$-radius achieved by using Babai's nearest plane algorithm using this basis $B=\lr{\bs{b}_1^T \dots \bs{b}_{2n}^T}^T  $ is given by 
\begin{equation}
    \epsilon=\frac{1}{2} \min_i \| \bs{\tilde{b}}_i \|,
\end{equation}
where we write $\Tilde{B}=({\bs{\tilde{b}}_1^{ T}, \dots ,\bs{\tilde{b}}_{2n}^{* T}})^T  $ for the Gram-Schmidt orthogonalization of $B$.

The secret key pair $(f, g)$ can be extended to a full rank secret basis \cite{NTRUSign, Ducas, StehleSteinfeld}
\begin{align}
    B_R= &\begin{pmatrix}
        f & g \\ F & G
    \end{pmatrix} \in R^{2 \times 2}\\
    &\mapsto
    B_{\Z}=\begin{pmatrix}
         C_{\Phi}\lr{f} & C_{\Phi}\lr{g} \\ C_{\Phi}\lr{F} & C_{\Phi}\lr{G}
    \end{pmatrix} \in \Z^{2n \times 2n},
\end{align}
which constitutes an equivalent lattice basis in $R^2$ if $F, G \in R$ are such that $fG-gF =q\in R$,  that is if 
\begin{align}
    B_RJB_R^T &=
   \begin{pmatrix}
        f & g \\ F & G
    \end{pmatrix}
   \begin{pmatrix}
        0 & 1 \\ -1 & 0
    \end{pmatrix}
       \begin{pmatrix}
        f & F \\ g & G
    \end{pmatrix}\nonumber \\
    &=
    \begin{pmatrix}
        0 & fG-gF \\ gF-fG & 0
    \end{pmatrix}=qJ
\end{align}
is symplectic in $R^2$.  The vector  $(F, G) \in R^2$ is typically chosen to be the minimal representative modulo multiples of $(f, g)$ in $R$ and can be approximated efficiently using Babai's algorithm \cite{NTRUSign, Ducas, StehleSteinfeld}.  By applying the circulant homomorphism,  this secret basis is mapped to a secret basis for the lattice in $\Z^{2n}$ on which the GKP-NTRU lattice is defined -- note that the symplectic basis obtained from the public key,  eq.~\eqref{eq:HCS} is obtained from this by a rotation and change of basis using $\sigma_{\Phi}$.  Due to this simple relationship,  we  perform the following analysis in the non-rotated basis w.l.o.g.~.

By leveraging symplecticity,  we can derive lower bounds on the \texttt{BDD}-radius achieved using this secret basis.  By comparing this to solving \texttt{BDD} when the input public basis is $\delta$-LLL reduced, we obtain an almost exponential separation between the BDD radius provided by the public and private basis.

\begin{prop}
    Let $q\geq n$.
    Using the secret basis 
    \begin{equation}    
    B=\begin{pmatrix}
        C_{\Phi}(f) & C_{\Phi}(g) \\ C_{\Phi}(F) & C_{\Phi}(G) 
    \end{pmatrix},
    \end{equation}
Babai's algorithm solves $\mathtt{BDD}_{\epsilon}$ with 
\begin{equation}
    \epsilon_B\geq q\lr{ 2\max\lrc{ \|\lr{\bs{f}, \bs{g}}\|, \|\lr{\bs{F}, \bs{G}}\|}}^{-1},
\end{equation}
which is at worst on a scale of $O\lr{1/\mathtt{poly}(n)}$,  while using the $\delta$-LLL reduced public basis obtained from
\begin{equation}
        H=\begin{pmatrix}
        I& C_{\Phi}(h) \\ 0 & qI 
    \end{pmatrix},
\end{equation}
we have
\begin{equation}
    \epsilon_H\leq \frac{\lambda_1\lr{L}}{2}e^{-\sqrt{n \ln{\lr{q}} \ln{\lr{1/\delta}}}},
\end{equation}
which is on a scale of $o({e^{-\sqrt{n\ln{n}\ln{\lr{1/\delta}}}}})$. 
\end{prop}

\proof

We define the conjugation of polynomials $R\mapsto R:\; f \mapsto \overline{f} \Leftrightarrow \bs{f}\mapsto \sigma_{\Phi}\bs{f}$ such that $C_{\Phi}(\overline{f})=C_{\Phi}(f)^T$.  Using the homomorphism to circulants we have that

\begin{align}
    B_{\Z}&=\begin{pmatrix}
        C_{\Phi}(f) & C_{\Phi}(g) \\ C_{\Phi}(F) & C_{\Phi}(G) 
    \end{pmatrix}, \\
    \overline{B}_{\Z}&=\begin{pmatrix}
        C_{\Phi}(\overline{f}) & C_{\Phi}(\overline{g}) \\ C_{\Phi}(\overline{F}) & C_{\Phi}(\overline{G}) 
    \end{pmatrix} , \\
    B_{\Z}J\lr{\frac{1}{q}\overline{B}_{\Z}}^T &= J,  \\
     \lr{\frac{1}{q} J^T\overline{B}_{\Z}}JB_{\Z}^T &=I ,
\end{align}
where the last line is obtained using the defining operation $fG-gF=q \in R$.  We will from now on omit the subscript $ _\Z$.
This equation identifies the canonical symplectic dual of the basis $B_{\Z}$ as
\begin{equation}
    B^{\perp}=\frac{1}{q} J^T\overline{B},
\end{equation}
which is related to the canonical euclidean dual by a $J-$rotation \cite{Conrad_2022}
\begin{equation}
    B^{*}=B^{\perp}J=\frac{1}{q} J^T\overline{B}J.
\end{equation}
Let $\Tilde{B}=\mathtt{GS}\lr{B}$ the Gram-Schmidt diagonalization of $B$ and \begin{equation}
    \Hat{B^*}=R_{2n}^T\mathtt{GS}\lr{R_{2n}B^*}=\frac{1}{q}R_{2n}^T\mathtt{GS}\lr{R_{2n}J^T\overline{B}}J,
\end{equation} the Gram-Schmidt diagonalization of the canonical euclidean dual $ B^{*}$ in reverse order,  where $ R_{i,j}=\delta_{n-i,j}$ and $R_{2n}=R\oplus R$.  We have that
\begin{equation}
    \|\bs{\tilde{b}}_i\|\|\bs{\hat{b^*}}_i\|=1\,\forall i,
\end{equation}
such that
\begin{equation}
    \min_i \|\bs{\tilde{b}}_i\|=\frac{1}{\max_i \|\bs{\hat{b^*}}_i\|}.
\end{equation}
The Gram-Schmidt norm is trivially upper bounded by
\begin{equation}
     \|\Hat{B^*}\|_{\rm GS}=\max_i \|\bs{\hat{b^*}}_i\| \leq \max\lrc{ \|\lr{\bs{f}, \bs{g}}\|, \|\lr{\bs{F}, \bs{G}}\|}/q,
\end{equation}
such that we obtain 
\begin{equation}
    \min_i \|\bs{\tilde{b}}_i\|\geq 1/\|\Hat{B^*}\|_{\rm GS}=q/\max\lrc{ \|\lr{\bs{f}, \bs{g}}\|, \|\lr{\bs{F}, \bs{G}}\|}.
\end{equation}
This lower bound on the Gram-Schmidt norm of the secret basis is large, when the secret key pairs $(f,g), (F,G)$,  with length on a scale of $\Omega\lr{\sqrt{n}}$ are short relative to $q \geq n$,  which is expected to be the case by construction of the cryptosystem. 
More concretely,  for $\Phi=x^n+1$,  the NTRUSign construction in \cite[Lemma 4.6]{StehleSteinfeld} asserts that $(F,G)$ can be found such that $\|(F,G)\| \leq \sigma n$,  where $\sigma \approx n^c \sqrt{q} $ is the standard deviation of the discrete Gaussian distribution used to sample the discrete Gaussians.
This bound yields a BDD radius not smaller than $\epsilon_B \sim O\lr{1/\mathtt{poly}(n)}$. 


We compare this to the bound obtained  from the public basis,  when the input basis $H$ is $\delta$-LLL reduced \cite{LLL}.  Building on an argument by Eldar and Hallgreen,  Ducas and van Woerden \cite{Eldar, DucasBDD} have shown that for a $q$-ary lattice (for $q=c^n$),  Babai's algorithm can solve BDD up to \begin{equation}
    \epsilon_H=\frac{\lambda_1\lr{L}}{2}e^{-\sqrt{n \ln{\lr{q}} \ln{\lr{1/\delta}}}},
\end{equation}
which has also been extended to general $q$ in ref.~\cite{AllenBDD}.   \endproof

\subsection{Quantum public key communication from NTRU-GKP codes}

In addition to its usual use as a QECC,  the fact that the NTRU-GKP codes have the additional property that \textit{decoding} for a stochastic displacement noise model is tightly related to \textit{decrypting} the NTRU cryptosystem suggests that the NTRU-GKP codes presented here may be used for both,  quantum error correction and a new kind of quantum public key communication scheme at the same time.  One may interpret NTRU-GKP codes as \textit{trapdoor decodable quantum error correcting codes}.  That is,  while stabilizer measurements can be performed and code states prepared using only access to the public key $h$, knowledge of the corresponding secret keys $\lr{f,  g}$ of the NTRU cryptosystem is necessary for reliable and efficient decoding.

In the following, we outline how instances of the NTRU-GKP code can be used to set up a \emph{private quantum channel} \cite{PrivateQuantumChannel} 
with quantum information being sent from Bob to Alice,  in that quantum information is transmitted in a fashion that is 
oblivious to an eavesdropper with limited computational power who has access to the physical quantum channel used for transmission. %
This setup is based on the observation that an attacker capable of decoding instances of the NTRU-GKP code by solving CVP on the related lattice also allows her successfully retrieve the message from the ciphertext of the corresponding NTRU cryptosystem.

The workings of the here proposed cryptosystem is similar to that of a \emph{one time pad} (OTP), where every OTP instance corresponds to a random displacement error applied to an NTRU-GKP code instance such that the syndrome of the random displacement error encodes a ciphertext of the NTRU scheme. The security of this scheme under the assumption that decoding a quantum error correcting code -- i.e.,  finding small errors that are consistent with the syndrome --  is \textit{necessary} to retrieve its logical content is then immediately inherited from the corresponding classical NTRU cryptosystem.  While we have carried out most of our exposition with the only heuristically secure version of the NTRU cryptosystem originally presented in  ref.~\cite{Hoffstein_Pipher_Silverman_1998} with quotient $\Phi_0=x^n-1$ and secret key sampling from the uniform binary distributions $D(d_1, d_2)\subseteq \lrc{-1,0, +1}^n$,  we have also shown that NTRU-GKP codes can be constructed using irreducible quotients $\Phi=x^n+1$,  $n$ a power of $2$,  and $q\geq \mathrm{poly}(n)$ while secret key pairs are sampled from discrete Gaussian distributions.  This is the setting chosen for a provably secure version of the NTRU cryptosystem discussed in ref.~\cite{StehleSteinfeld},  where the public key is shown to be pseudorandom and security is inherited from the (average-case) hardness of the ring based $\mathtt{R-SIS}$ and $\mathtt{R-LWE}$ problem.

\begin{figure*}
     \centering
\includegraphics[width=\textwidth]{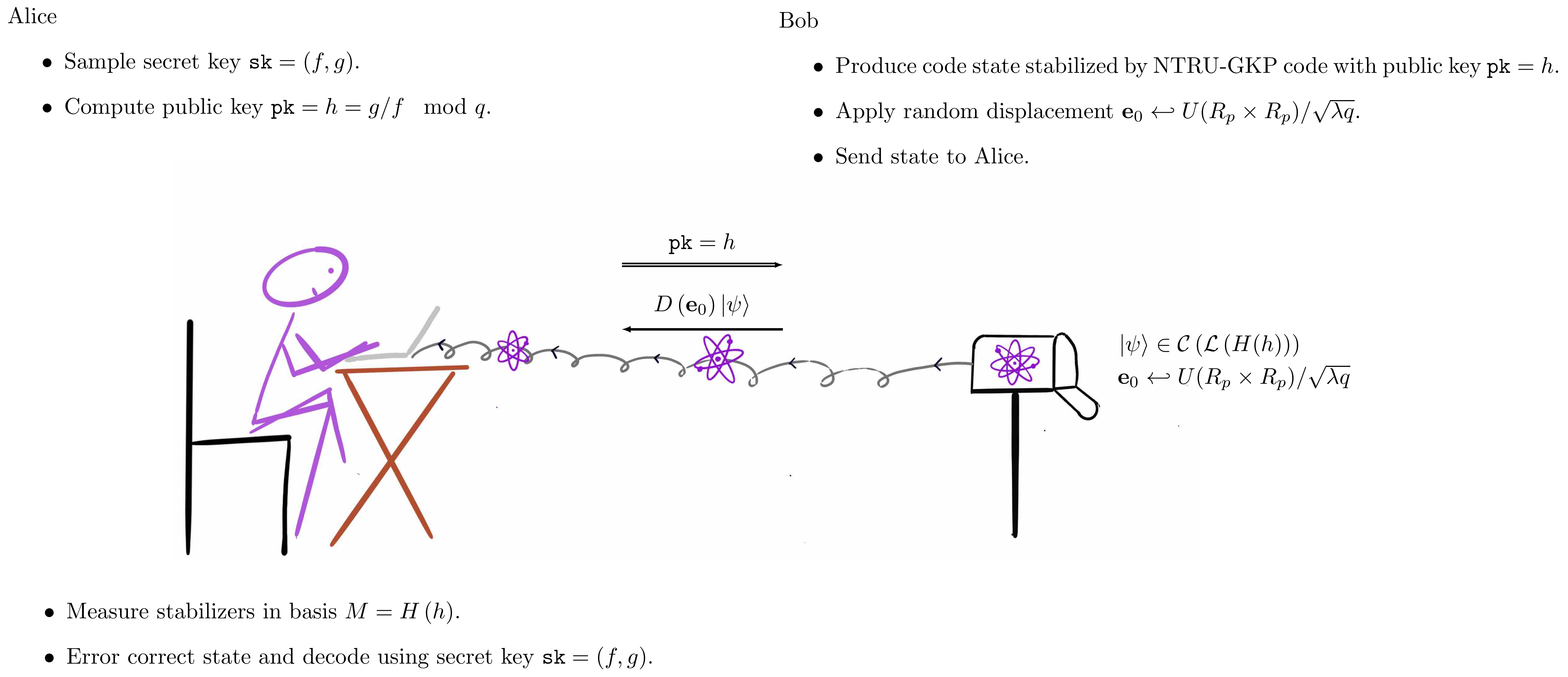}
     \caption{Outline of the private quantum channel established using the NTRU-GKP code as described in the main text. 
     }
     \label{fig:PQC}
 \end{figure*}

The public key protocol,  also described in fig.~\ref{fig:PQC} is sketched as follows:
\begin{enumerate}
\item Alice samples a secret key pair $(f,g)$ and computes the public key $h$,  which is communicated to Bob.
\item Bob produces a code state described by the GKP code using the basis $\sqrt{({\lambda}/{q})} H(h)$ and samples an error corresponding to a random message $\bs{e}_0=(-\bs{r}, \bs{m})/\sqrt{\lambda q}$,  according to the specifications of the NTRU cryptosystem,  by which he displaces the state. He transmits the state to Alice.

\item Alice measures the stabilizers and decodes the state, e.g., via the NTRU decryption routine or by employing Babai's algorithm as outlined before using the secret key pair $(f,g)$.  She has hence received the to her unknown state from Bob through the error corrected private quantum channel. 

\end{enumerate}

To our knowledge, this setup presents a new paradigm of quantum cryptographic protocols.  We summarize points in support of its security.

\paragraph{Necessity to decode.} In order to unambiguously obtain the logical code state,  it is necessary to find a correction $\bs{e'}$ consistent with the syndrome such that $ \|\bs{e}_0+\bs{e'}\| \leq \Delta /2$.  Since $\|\bs{e}_0\|_{\infty} \leq 1/\sqrt{\lambda q}$ and the smallest element in $\CL^{\perp}$ is of length $\Delta$,  this amounts to decrypting the NTRU ciphertext in the syndrome to identify $\bs{e}_0$.  Towards a first cryptanalysis,  we examine to which degree an adversary is able to distinguish a quantum ciphertext.
Let $\ket{\overline{\psi}}$ be a logical code state vector specified by a GKP-NTRU code with lattice $\CL$.  We examine the eigenvalue of logical Pauli observables
obtained when the initial code state is encrypted by applying the random displacement $D\lr{\bs{e}_0}$,  a syndrome $\bs{s}\lr{e_0}=MJ\bs{e}_0 \mod 1$ is obtained and a generic correction via $\bs{\eta}=(MJ)^{-1}\bs{s}\lr{\bs{e}_0}$ is applied. 
With 
\begin{align}
M^{-1}=\frac{1}{\sqrt{\lambda q}}
\begin{pmatrix}
    qI & -A_{\Phi}\lr{h} \\ 0 & I
\end{pmatrix},
\end{align}
this yields a generic correction
\begin{align}
    \bs{\eta}=\frac{1}{\sqrt{\lambda q}} 
    \begin{pmatrix}
        0 \\ \bs{c},
    \end{pmatrix}
\end{align}
where $\bs{c}=\bs{m}+A_{\Phi}(h)\bs{r} \mod q$ is the associated NTRU ciphertext. 
The total remaining error after correction thus is 
\begin{align}
\bs{e}_0-\bs{\eta}=\frac{-1}{\sqrt{\lambda q}}
    \begin{pmatrix}
         \bs{r} \\ \bs{c}-\bs{m}
    \end{pmatrix}=    \begin{pmatrix}
         \bs{r} \\ A_{\Phi}(h)\bs{r} \mod q
    \end{pmatrix}.
\end{align}
We compute
\begin{equation}
M^{\perp}J\lr{\bs{e}_0-\bs{\eta}}=\frac{1}{\lambda}\begin{pmatrix}
0 \\ \bs{r}  
\end{pmatrix} \mod q/\lambda,
\end{equation}
which shows that for an input code state vector $\ket{\overline{\psi}}$,  after encoding and generic correction,  the eigenvalues of logical Pauli operators corresponding to rows $i=n+1,\dots, 2n$ in $M^{\perp} $ obtain a random phase $e^{i\frac{2\pi}{\lambda}r_{i-n}}$.  This observation suggests that,  for $\lambda=2$,  without access to the random string $\bs{r}$ embedded in the NTRU ciphertext in every instance,  the quantum state is effectively projected onto a state that is diagonal in the logical Pauli-$Z$ basis and quantum superpositions are washed out.  This situation is similar to that of half a quantum OTP,  where only one type (either $X$ or $Z$) of Pauli operators is used in the encryption. 

\paragraph{Orthogonality.} For a fixed quantum state vector $\ket{\overline{\psi}}$,  different error realizations $D\lr{\bs{e}_0}$ where $\|\bs{e}_0\|< \Delta /2$ map the state to mutually orthogonal states (sectors of the QECC).  This is guaranteed by the quantum error correction conditions.  Without applying suitable corrections,  separate encodings of the same logical quantum state vector $D\lr{\bs{e}_i}\ket{\overline{\psi}}$ are expected to appear uncorrelated.

We leave a further study of the degree of quantum security of this scheme as challenge for future work.  It is important to stress that the security of this scheme is not based on information theoretic arguments,  but on computational limitations of an eavesdropper,  giving rise to a new situation in quantum cryptography. 
For a practical security analysis it would be furthermore meaningful to study how potential security claims can sustain when also considering the finite squeezing error present in physical GKP states \cite{Terhal_2020, Tzitrin_2020}.

In addition to the possibility of obtaining a classical public key private quantum channel by this construction,  this scheme is also expected to be tolerant against additional errors imposed by the channel. Additional displacement errors would effectively change the ``quantum encoded" NTRU ciphertext as encoded in the syndrome, but as long as the additional error together with the initial random displacement implemented by Bob are sufficiently smaller than the euclidean code distance, the transmitted logical quantum state is still expected to be decoded correctly.

As computation and transmission of quantum information encoded in GKP codes using photonics and integrated optics is becoming technologically ever more developed, this setup is interesting for the reason that to transmit quantum information one would potentially use  a (bosonic) quantum error correcting codes anyways. 
 Our construction highlights that this can be done with in-built security options without explicitly concatenating the encoded qubits into a separate cryptographic protocol.

\section{Conclusion and outlook}\label{sec:conclusion}
In this work, we have 
introduced the randomized construction of \textit{good} GKP codes using the NTRU cryptosystem and discussed how a decoder for these codes can be obtained from variations of the NTRU decryption process. 

 We have discussed the use of these codes in a public key quantum communication scheme where we expect to inherit a conditional security guarantee from the original cryptosystem. This defines a \textit{trapdoor decodable quantum error correcting code} for which the core idea is that we can provide ``bad" bases for suitably chosen GKP codes that allow an agent to prepare code states and measure stabilizers but -- without access to a ``good" secret basis -- require exponential overhead to decode the syndrome. We leave as open challenge to either prove or disprove the quantum security of our scheme.
More broadly,  this idea also opens the door to potential client-server schemes where a client requests a server -- capable of preparing GKP states and carry out Gaussian operations -- to carry out computations on client-specified GKP codes and measure stabilizer syndromes without giving the server the power to decode efficiently to apply logical corrections.  

 It is also worth mentioning that the NTRU cryptosystem has multi-key homomorphic properties with respect to adding and multiplying the message- and random bit-strings \cite{NTRUHomo}.  Beyond the scope of this work, we expect it to be possible to leverage these homomorphic properties to design more advanced NTRU-GKP codes for the secure and error corrected transmission of quantum states.  Effective NTRU-lattices derived from products of public/secret keys correspond to non-principal ideals of the underlying ring,  which makes for an interesting generalization of our setting. Alternatively,  it would also be interesting to examine symplecticity for higher rank module lattices.  Higher rank module lattices over polynomial rings have previously found application in quantum error correction, e.g.,  in the work of Pantaleev and Kalachev \cite{Pantaleev_2022} to construct high distance quantum LDPC codes,  who consider binary polynomial rings with quotient $\Phi_0=x^l-1$ and where the element-wise homomorphism of basis elements of the $R$-module basis to circulants is termed \textit{lift}.

On the \emph{hardware level},  we also expect that the cyclic structure of these codes can be helpful in the design of modular stabilizer-measurement architectures with a fixed stabilizer-measurement gadget that is coupled to the data modes with shifted mode-index and at alternating delay.  The short length of the corresponding displacements when the stabilizers generators are measured in the secret basis implies a reduced overhead in required interaction time/strength when the required connectivity is present and further highlights a \textit{physical} advantage in having access to the secret basis for several meaningful physical platforms.

For future work,  it would be interesting to improve on the decoders,  such as adapting our proposition of \texttt{BabaiDecode} by adapting Babai's nearest plane algorithm to include information about the biased input error distribution eq.~\eqref{eq:res_prob},  to design a better approximation to MLD decoding NTRU-GKP codes and provide numerical studies for large $n$.

We have highlighted the complexity of decoding GKP codes as an interesting subject to study and we expect that,  using concatenation,  computational complexity questions on the GKP-lattice level can be put into tighter relationships with corresponding problems in qubit-codes and equivalent questions in classical error correction.  It would further be interesting to identify other ideal lattices that can be used to construct GKP codes.

Finally,  it is worth mentioning that the relationship between GKP quantum error correction and cryptography runs even deeper.  Physical,  i.e.,  normalizable,  realizations of GKP states obey a phase-space probability distribution very similar to that of a discrete lattice Gaussian distribution.  Quantum states as such,  and the ability to produce and sample from them,  play a central role in the quantum reduction from SVP to the learning with errors problem \cite{Regev_2005}.  Given the ability to efficiently prepare approximate GKP-state by measuring its stabilizers,  we leave as final open question in how far it is possible to sample from discrete lattice Gaussian distributions using physically preparable GKP states.

\acknowledgements 
We thank 
Victor V.~Albert, 
Yusuf Alnawakhtha, 
Francesco Arzani, 
Nikolas P.~Breuckmann, 
Steven T.~Flammia, 
Cica Gustiani, 
Florian Hirsch, 
John Preskill,
Manasi Shingane, 
Vincent Ulitzsch, and 
Daochen Wang for many helpful and inspiring discussions.  We thank Henning Seidler for constructive feedback on parts of an early version of the manuscript and Nathan Walk for many helpful comments on a later version.  JC thanks the Joint Center for Quantum Information and Computer Science (QuICS) at the University of Maryland for their kind hospitality during the final preparation stages of this project and many stimulating discussions.  During preparation of this manuscript JC was also temporarily affiliated with the AWS CQC and Caltech,  which he also thanks for their hospitality.

We gratefully acknowledge support from the BMBF (RealistiQ, MUNIQC-Atoms, PhoQuant, QPIC-1, and QSolid,  6G-RIC, Q-Fiber, Q-net-Q),  the DFG (CRC 183, project B04, on entangled states of matter), 
the Quantum Flagship (Millenion, PasQuans2), 
the Munich Quantum Valley (K-8), the ERC (DebuQC), the EU (Q-net-Q)
as well as the Einstein Research Unit on quantum devices,  for which this is an inter-node joint project as well as Berlin Quantum.

\bibliographystyle{quantum}
\bibliography{blindgkp_bib}
\onecolumn\vfill

\appendix

\section{$\lambda_1 = \Omega\lr{\sqrt{n}}$ scaling for NTRU-like lattices with random symmetric $H$.}\label{appendix:goodrandomH}

In this section we discuss a strategy towards proving conjecture \ref{conj:random_h} and \ref{conj:SS}.  Following the proof strategy laid out in ref.~\cite{Sarnak1994},  we show that a certain distribution over symplectic lattices with NTRU lattice-like generating matrix implement the ``design property" of eq.\  \eqref{eq:lattice_design},  which,  following the derivation in the main text,  suffices to establish the scaling provided by the Gaussian heuristic and thus establish goodness of the associated GKP codes.  By viewing the considered GKP codes as concatenated codes with a $\CL_{\square}\propto q\Z^{2n}$ sublattice,  the following statement also establishes the existence of ``good" qudit-based quantum error correcting codes when the local dimension $q\rightarrow\infty$ is large and yields an alternative proof of the existence of good qudit-based stabilizer codes as similarly obtained from the quantum Gilbert-Varshamov bound \cite{gottesman_thesis, ashikhmin1999quantum, ashikhmin2000nonbinary}. The proof presented here also yields a simple constructive way to sample GKP- or qudit-based stabilizer codes that are expected to be good.
A canonical basis for an NTRU lattice is given by the rows of the $ \Z^{2n \times 2n}$ matrix 

\begin{equation}
    M\lrq{X}=\begin{pmatrix}
        I & X \\ 0 & qI_n
    \end{pmatrix},
\end{equation}
where $X=H(h) \in \F_q^{n \times n}$ has a special cyclic structure.  The top block can be interpreted as the reduced row-echelon form $\lr{I \; H\lr{h}}$ of a classical linear $q-$ary code in $\mathbb{F}_q^{2n}$.
For $X=X^T$ symmetric, we have that $M\lrq{X}$ is $q-$ symplectic. We will denote $M_{\rm sp}\lrq{X}=M\lrq{X}/\sqrt{q}$ its rescaling to a symplectic matrix.
Following the technique used in ref.\ \cite{Sarnak1994},  we first show the subsequent statement.

\begin{prop}\label{prop:randsymH}
Let
\begin{equation}
U_q := \lrc{X=X^T \in \lrc{-\frac{q}{2} ,\hdots , \frac{q}{2} }^{n \times n} }
\end{equation}
be the set of symmetric matrices in $\Z_q$ and let $f:\, \R^{2n} \rightarrow \R$ be a function with compact support.  We have

\begin{equation}
  \lim_{q\rightarrow\infty}  \Big\langle  \sum_{\bs{k}\in \Z^{2n}-\lrc{0}}  f\lr{M^T_{\rm sp}\lrq{X} \bs{k}} \Big\rangle_{X \in U_q} = \int_{\R^{2n}} f\lr{\bs{x}} d\bs{x},
\end{equation}
where the expectation value on the LHS is taken uniformly over $U_q$.

\end{prop}

\proof
We start from the definition
\begin{align}    
\lim_{q\rightarrow \infty}\Big\langle F(X)\Big\rangle_{X \in U_q} &= \lim_{q\rightarrow \infty} q^{-1}\sum_{X_{1,1} = -q/2}^{q/2} q^{-1}\sum_{X_{1,2} =-q/2}^{q/2} \hdots F(X)  \\
&= \int_{-1/2}^{1/2} dX_{1,1} dX_{1,2}dX_{1,3}\hdots F(qX) 
\nonumber.
\end{align}
%
We have for $\bs{k}=\bs{m} \oplus \bs{n}$
\begin{align*}
M^T_{\rm sp}\lrq{X} \bs{k} = q^{-\frac{1}{2}}\begin{pmatrix}
    \bs{m} \\ qX\bs{m} +q\bs{n} 
\end{pmatrix},    
\end{align*}
such that we can compute analogously to
the argument presented in ref.\ \cite{Sarnak1994}
\begin{align}
I(q)&=\int_{-1/2}^{1/2} dX_{1,1} dX_{1,2}dX_{1,3}\hdots \sum_{\bs{m},\bs{n} \in \Z^{n}-\lrc{0}}  f\lr{q^{-\frac{1}{2}}\begin{pmatrix}
    \bs{m} \\ qX\bs{m} +q\bs{n} 
\end{pmatrix}} \label{eq:sum_dec0}\\
&=\int_{-1/2}^{1/2} dX_{1,1} dX_{1,2}dX_{1,3}\hdots \lrc{ \sum_{\substack{ \bs{m} \in Z^n,\\ m_1\neq 0 }}+\sum_{\substack{ \bs{m} \in Z^n,\\ m_1=0 \\ m_2 \neq 0 }} + 
\sum_{\substack{ \bs{m} \in Z^n,\\ m_1=0 \\ m_2 =0 \\ m_3 \neq 0 }} + \hdots
}\sum_{\bs{n} \in \Z^n} f\lr{q^{-\frac{1}{2}}
\begin{pmatrix}
    \bs{m} \\ qX\bs{m} +q\bs{n} 
\end{pmatrix}} \label{eq:sum_dec} \\ 
&+\sum_{\bs{n} \in \Z^n} f\lr{
\begin{pmatrix}
    \bs{0} \\ \sqrt{q}\bs{n} 
\end{pmatrix}} \label{eq:sum_dec2}.
\end{align}
In eq.\  \eqref{eq:sum_dec}, we consider each summation over $\bs{m}$ separately. 
In the first term with the constraint $m_1 \neq 0$
we rewrite
\begin{align}
    qX\bs{m}+q\bs{n}=\begin{pmatrix}qm_1\lr{X_{1,1}+\frac{n_1}{m_1} + m_1^{-1}\sum_{k>1}X_{1,k} m_k  }
    \\ 
    qm_1\lr{X_{2,1}+\frac{n_2}{m_1} + m_1^{-1}\sum_{k>1}X_{2,k} m_k  }
    \\ 
    qm_1\lr{X_{3,1}+\frac{n_3}{m_1} + m_1^{-1}\sum_{k>1}X_{3,k} m_k  } \\
    \vdots
    \end{pmatrix}\label{eq:sum_vec}.
\end{align}
We write for each $n_i=\lfloor \frac{n_i}{m_1}\rfloor m_1 + (n_i \mod m_1)$ and split the summation
\begin{equation}
    \sum_{n_i\in \Z} g\lr{\frac{n_i}{m_1}} = \sum_{j_i\in \Z}\sum_{n_i\in \Z_{m_1}} g\lr{j_i+ \frac{n_i}{m_1}}.
\end{equation}
This way,  each summation over the integer divisors of $n_i$ with $m_1$ can be combined with the integral over $X_{i,1} \in \lrq{-1/2,1/2}$ to an integral of $X_{i,1}  \in \R$ over the real numbers.  To perform this trick,  start with $X_{1,1} + j_1$ in the first row of eq.\  \eqref{eq:sum_vec} and realize that all subsequent rows are independent of $X_{1,1}$.  After converting the integration in the first row, all remaining summand of that row can be absorbed into a shift of the $X_{1,1}$ integral.
Now the first row is also independent of $X_{2,1}=X_{1,2}$,  such that we can repeat this trick,  converting 
the integral over $X_{2,1}$  and summation over $j_2$ into integration of $X_{2,1}$ over $\R$ which again gets rid of the dependency on $X_{k,2}, k>1$ in this row. 
Similarly,  the summations over the terms $\frac{n_i}{m_1}$ also becomes trivial and provides a factor of $m_1$.  In total, after substitution $t_i=qm_1X_{i,1}$
\begin{align}
&\int_{-1/2}^{1/2} dX_{1,1} dX_{1,2}dX_{1,3}\hdots \sum_{\substack{ \bs{m} \in Z^n,\\ m_1\neq 0 }}
\sum_{\bs{n} \in \Z^n} f\lr{q^{-\frac{1}{2}}
\begin{pmatrix}
    \bs{m} \\ qX\bs{m} +q\bs{n} 
\end{pmatrix}}\\
\nonumber
&=\sum_{\substack{ \bs{m} \in Z^n,\\ m_1\neq 0 }}
\int_{-\infty}^{\infty} d\bs{t}\,
q^{-n}f\lr{q^{-\frac{1}{2}}
\begin{pmatrix}
    \bs{m} \\ \bs{t}
\end{pmatrix}}\\
&=q^{-n/2}\sum_{\substack{ \bs{m} \in Z^n,\\ m_1\neq 0 }}
\int_{-\infty}^{\infty} d\bs{t}\,
f\lr{
\begin{pmatrix}
    q^{-\frac{1}{2}}\bs{m} \\ \bs{t}
\end{pmatrix}
}.
\nonumber
\end{align}
In the second term with constraint $m_1=0,\, m_2\neq 0$ we repeat the above procedure by pulling out a factor of $qm_2$, $qX\bs{m}+q\bs{n}=qm_2( qX\bs{m}/m_2 + \bs{n}/m_2)$.  Begin with the integration over $X_{2,2}$,  together with the sum over $n_2$ this again extends the domain of integration of $X_{2,2}$ to $\R$.  Substituting the remaining summands in the corresponding row renders the rest of $qX\bs{m}+q\bs{n}$ independent of independent of $X_{2,i}, i>2$ such that in each other row we can combine the $X_{2,i}$ integration with the sum over $n_i$ to extend the domains of integration. 
Repeat this procedure using each $m_i\neq 0$ in eq.\ ~\eqref{eq:sum_dec} and finally use that $f$ has compact support, such that in the limit $q\rightarrow\infty$
eq.\ ~\eqref{eq:sum_dec0} becomes

\begin{equation}
   \lim_{q\rightarrow \infty} I(q)= \lim_{q\rightarrow \infty} q^{-n/2}\sum_{\substack{ \bs{m} \in Z^n-\lrc{0} }}
\int_{-\infty}^{\infty} d\bs{t}\,
f\lr{
\begin{pmatrix}
    q^{-\frac{1}{2}}\bs{m} \\ \bs{t}
\end{pmatrix}
}.
\end{equation}
In the limit, we again use the definition of the Riemann integral to finally obtain
\begin{equation}
   \lim_{q\rightarrow \infty} I(q)=\int_{\R^{2n}} f\lr{\bs{x}} d\bs{x}.
\end{equation}

\endproof

We make the observation that for this proof strategy to work, it was important that each row/column of $X$ contained one entry that was independent of all the other rows. This is manifestly not the case for the (quasi) cyclic matrices $H(h)$ provided by the NTRU cryptosystem.  To show goodness for NTRU-GKP codes for random public key $h$ and thus establish conjecture \ref{conj:random_h},  it would further be necessary to understand how the distribution over cyclic matrices $H(h)$ approximates that of the random symmetric matrices considered above.  We leave this as interesting open problem to examine in future work.

\section{Thresholds of GKP codes} \label{appendix:threshold}

In this section, we sketch how the existence and value of a threshold for a GKP code family can be analyzed using the lattice theta function.
For simplicity,  assume a zero syndrome $\bs{s}=0$ on a lattice $\mathcal{L}_n$ that is part of a family of lattices scaling with $n$.  The probability for the state to be in the $\bs{\xi}^{\perp}_n=0$ coset is given by $P([\bs{\xi}^{\perp}_n] | 0)= \Theta_{\mathcal{L}_n}(z) / \Theta_{\mathcal{L}^{\perp}_n} (z)$,  where 
\begin{equation}
\Theta_{\mathcal{L}^{\perp}_n}(z) = \sum_{\bs{\xi}^{\perp}_n \in \mathcal{L}^{\perp}_n/\mathcal{L}_n } \Theta_{\mathcal{L}_n+\bs{\xi}^{\perp}_n}(z) 
\end{equation}
denotes the probability to be in any logical coset and $z={i}/({2\pi \sigma^2})$.  A necessary condition for a GKP code family to exhibit a threshold is satisfied if there exists $z^* \in i \mathbb{R}$ such that for  any $|z|>|z^*|, \, z \in i \mathbb{R}$, it holds that
\begin{equation}
0=\lim_{n \rightarrow \infty}  \Theta_{\mathcal{L}^{\perp}_n}(z)- \Theta_{\mathcal{L}_n}(z) =\lim_{n \rightarrow \infty}  \sum_{ \bs{\xi}^{\perp}_n \in \mathcal{L}^{\perp}_n\setminus \mathcal{L}_n } \Theta_{\mathcal{L}_n+\bs{\xi}^{\perp}_n}(z).
\end{equation}
 We write 
 \begin{align}
 \Theta_{\mathcal{L}_n+\bs{\xi}^{\perp}_n}(z)
 &=\sum_{\delta \in \mathcal{D}_n} N_{\delta}(\mathcal{L}_n, \bs{\xi}^{\perp}_n) q^{\delta} \\
&= \sum_{\delta \in \mathcal{D}_n} \exp{ \left\{\delta \left(\delta^{-1}\ln(N_{\delta}(\mathcal{L}_n, \bs{\xi}^{\perp}_n) )-  |\ln(q)| \right) \right\}},\nonumber
 \end{align}
 where we have $\mathcal{D}_n:=\{\|\bs{x}\|^2,\; \bs{x} \in \mathcal{L}_n+\bs{\xi}^{\perp}_n) \}$ and $N_{\delta}(\mathcal{L}_n, \bs{\xi}^{\perp}_n)=\#\{\bs{x} \in \mathcal{L}_n+\bs{\xi}^{\perp}_n:\, \|\bs{x}\|^2=\delta\}$.  
 We have that $\lambda_1\lr{\mathcal{L}_n +\bs{\xi}^{\perp}_n}$ corresponds to a shortest representative of the logical coset given by $\xi_n^{\perp}$,  such that the threshold condition becomes
 \begin{align}
 0&=\lim_{n \rightarrow \infty}  \sum_{ \xi^{\perp}_n \in \mathcal{L}^{\perp}_n\setminus \mathcal{L}_n } \sum_{\delta \in \mathcal{D}_n} \exp{ \left\{\delta \left(\delta^{-1}\ln(N_{\delta}(\mathcal{L}_n, \bs{\xi}^{\perp}_n))-  |\ln(q)| \right) \right\}} \\
 &=\lim_{n \rightarrow \infty}  \exp{ \left\{\Delta^2 \left(\Delta^{-2}\ln(N_{\Delta^2}(\mathcal{L}_n, \bs{\xi}^{\perp}_n) )-  |\ln(q)| \right) \right\}} + ...,\nonumber
 \end{align}
 where $\Delta$ is the Euclidean code distance of the code $\mathcal{L}_n$ as defined in previous sections that we assume to grow with increasing $n$.  Since each term in the sum is positive,  a necessary condition for asymptotic error suppression becomes
 \begin{align}
 |\ln(q)|&>\Delta^{-2}\ln(N_{\Delta^2}(\mathcal{L}_n, \bs{\xi}^{\perp}_n) ) \\
 \Leftrightarrow 2\sigma^{2} &<\Delta^{2}\ln^{-1}(N_{\Delta^2}(\mathcal{L}_n, \bs{\xi}^{\perp}_n) ).
 \nonumber
 \end{align}
 Hence,  under negligence of all higher order terms we can upper bound the threshold as
 \begin{equation}
 2\sigma^{*\,2} < \min_{ \bs{\xi}^{\perp}_n \in \mathcal{L}^{\perp}_n /\mathcal{L}_n } \frac{\Delta^{2}}{\ln\lr{N_{\Delta^2}(\mathcal{L}_n, \bs{\xi}^{\perp}_n)}},
 \end{equation}
 which shows the impact of the entropic contribution $N_{\Delta^2}(\mathcal{L}_n, \bs{\xi}^{\perp}_n)$ on the potential threshold.

\section{Numerical results on decoding NTRU-HPS}\label{app:numerics}

In this section we report some small scale numerical experiments we have conducted on decoding the NTRU-GKP code (with $\Phi_0=x^n-1$) in the small $n$ regime,  where numerical experiments where feasible within the scope of this work. 
The following codes are obtained as the NTRU lattice with the largest shortest vector length amongst  $100$ samples of NTRU key pairs $(f,g)$, where each instance of the SVP problem is solved by full HKZ reduction.  We aim at correcting errors up to a standard deviation $\overline{\sigma}^*=\sigma^* /\sqrt{2\pi}$ with physical standard deviation of $\sigma^*=0.1$.  By solving for $n,q $ in $\Delta /2 \geq \sqrt{2n\overline{\sigma}^2}$ using the bound in proposition \ref{prop_1}, we trial $q=8$ as a reasonable parameter.

The parameters of the codes we obtained are summarized in 
fig.~\ref{fig:params}.  For comparison,  notice that a standard square GKP code concatenated with a small $[\![n,1,3 ]\!]$ qubit code has distance $\Delta=\sqrt{3/2}=1.22$ using typically $n=5$ to $9$ qubits to encode a single logical qubit while here a similar distance is achieved while encoding $k=n$ logical qubits.

\begin{figure}
\centering
         \begin{tabular}{c|c| c }
           $n,d,q,p$    &  $\lambda_1\lr{L_{\rm cs}}$   & $\Delta$      \\ \hline
           $7,2,8,3$    & $4$                       & $1$               \\
           $11,3,8,3$   & $4.69$                    & $1.17$             \\
           $17,5,8,3$   & $4.9$                     &$1.23$             \\
           $23, 7,8,3$  & $5.48$                    & $1.37$            
        \end{tabular}
        \caption{Parameters of sampled NTRU lattices.}\label{fig:params}
\end{figure}

\begin{figure*}
  \includegraphics[width=\textwidth]{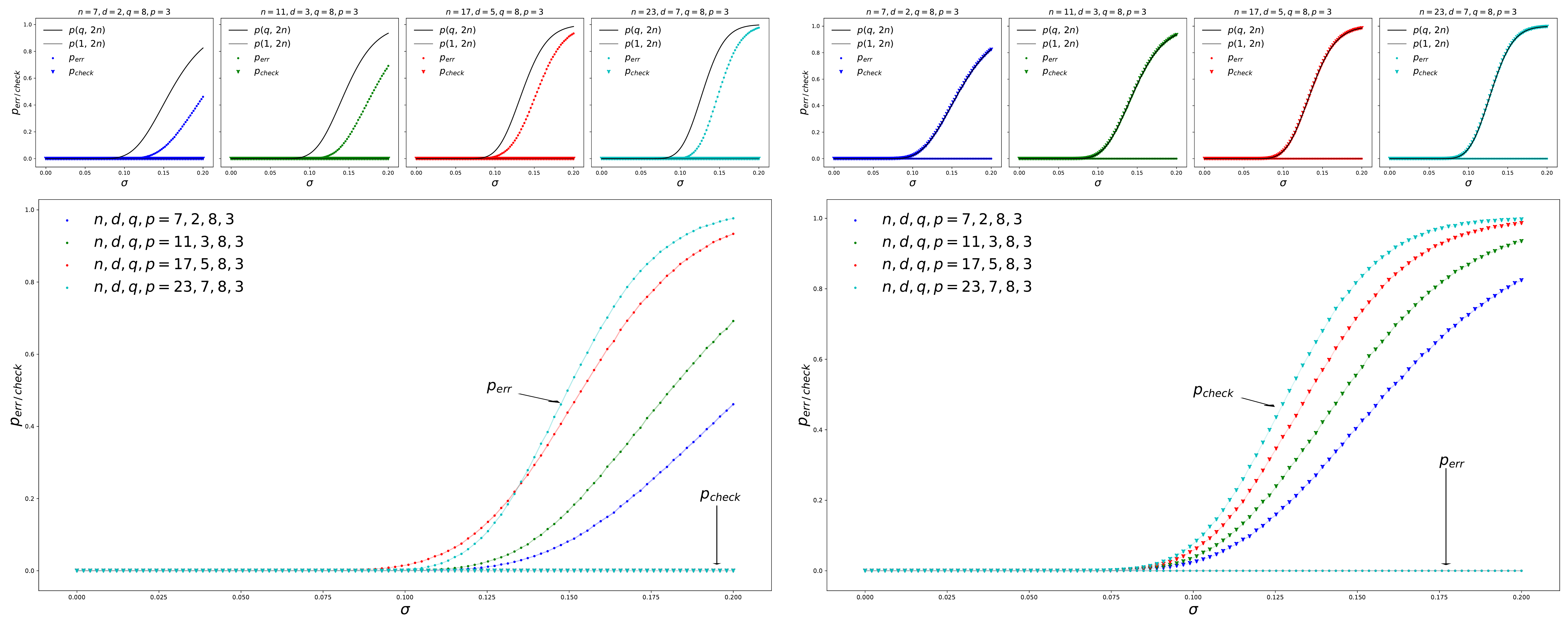}
\caption{Numerical results for the NTRU-GKP codes using the NTRU decryption routine \texttt{NTRUDecode} (left) and \texttt{BabaiDecode} (right) for NTRU-GKP lattices where $h$ is invertible.  Here,  the parameters $q=8$ and $p=3$ are fixed. $p_{\rm err}$ (dots) denotes the the logical error rate conditioned on successful decoding and $p_{\rm check}$ (stars) denotes the rate of decoding failures. }\label{fig:NTRU_dec_q8}
\end{figure*}

\begin{figure}
   \includegraphics[width=\textwidth]{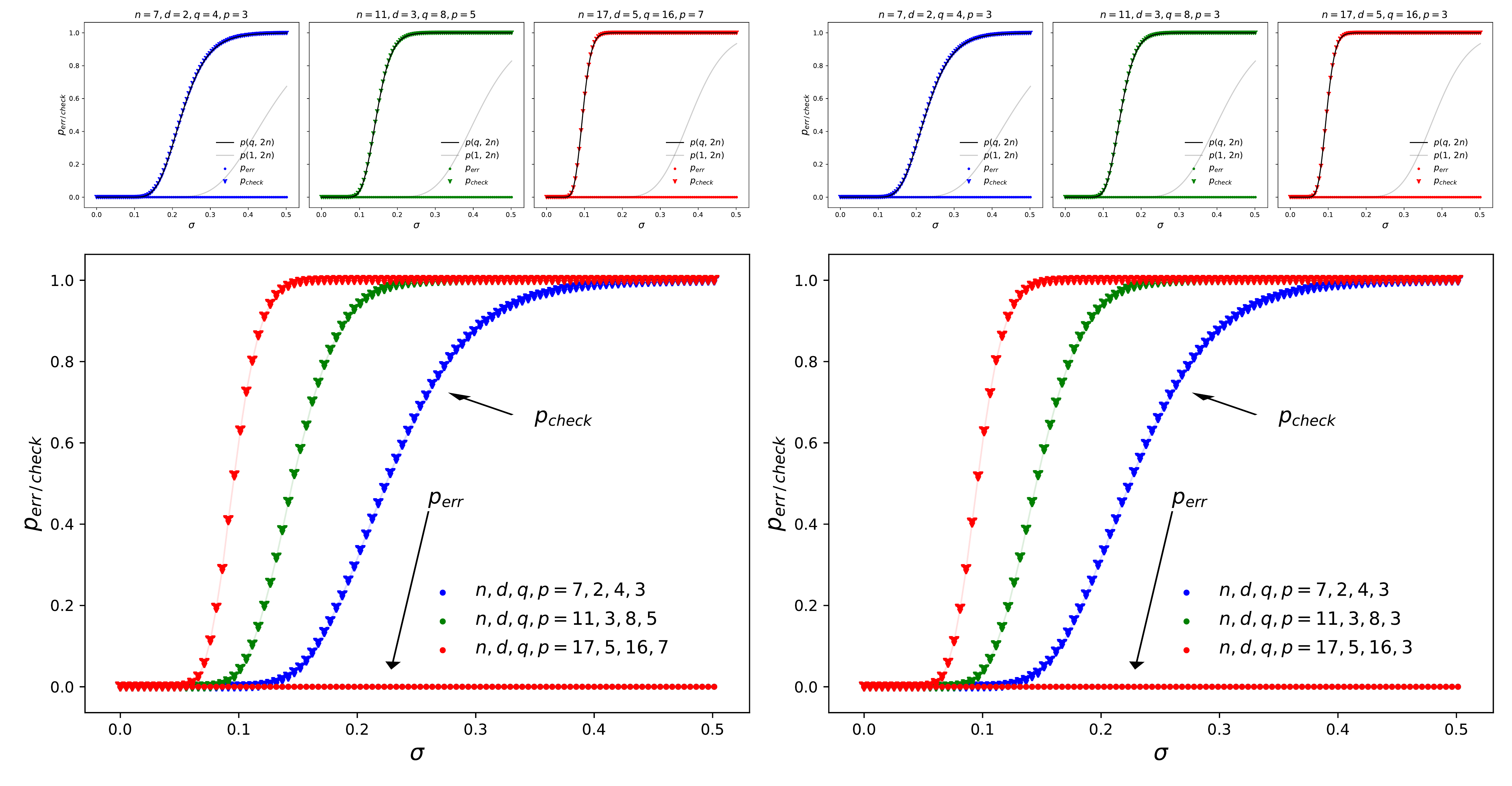}
 \caption{Numerical results for the NTRU-GKP codes using the NTRU decryption routine \texttt{BabaiDecode}. (left) $p=3,5,7$ is running and (right) $p=3$ is fixed.}\label{fig:BAB_dec}
 \end{figure}

\begin{figure}
   \includegraphics[width=\textwidth]{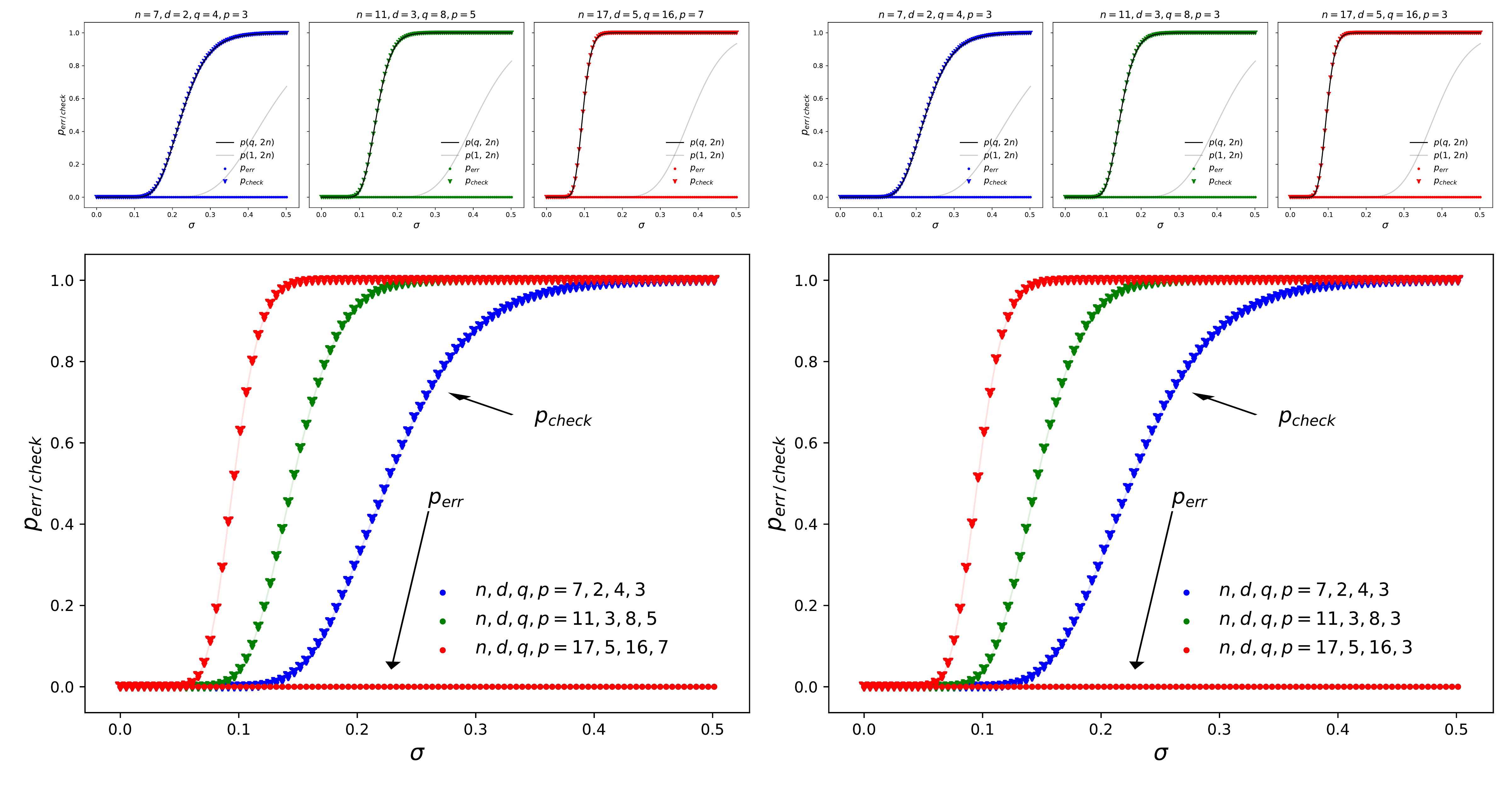}
 \caption{ Numerical results for the NTRU-GKP codes using the NTRU decryption routine \texttt{BabaiDecode} for NTRU-GKP codes where $h$ is invertible. (left) $p=3,5,7$ is running and (right) $p=3$ is fixed.} \label{fig:BAB_dec_hinv}
 \end{figure}

\begin{figure}
   \includegraphics[width=\textwidth]{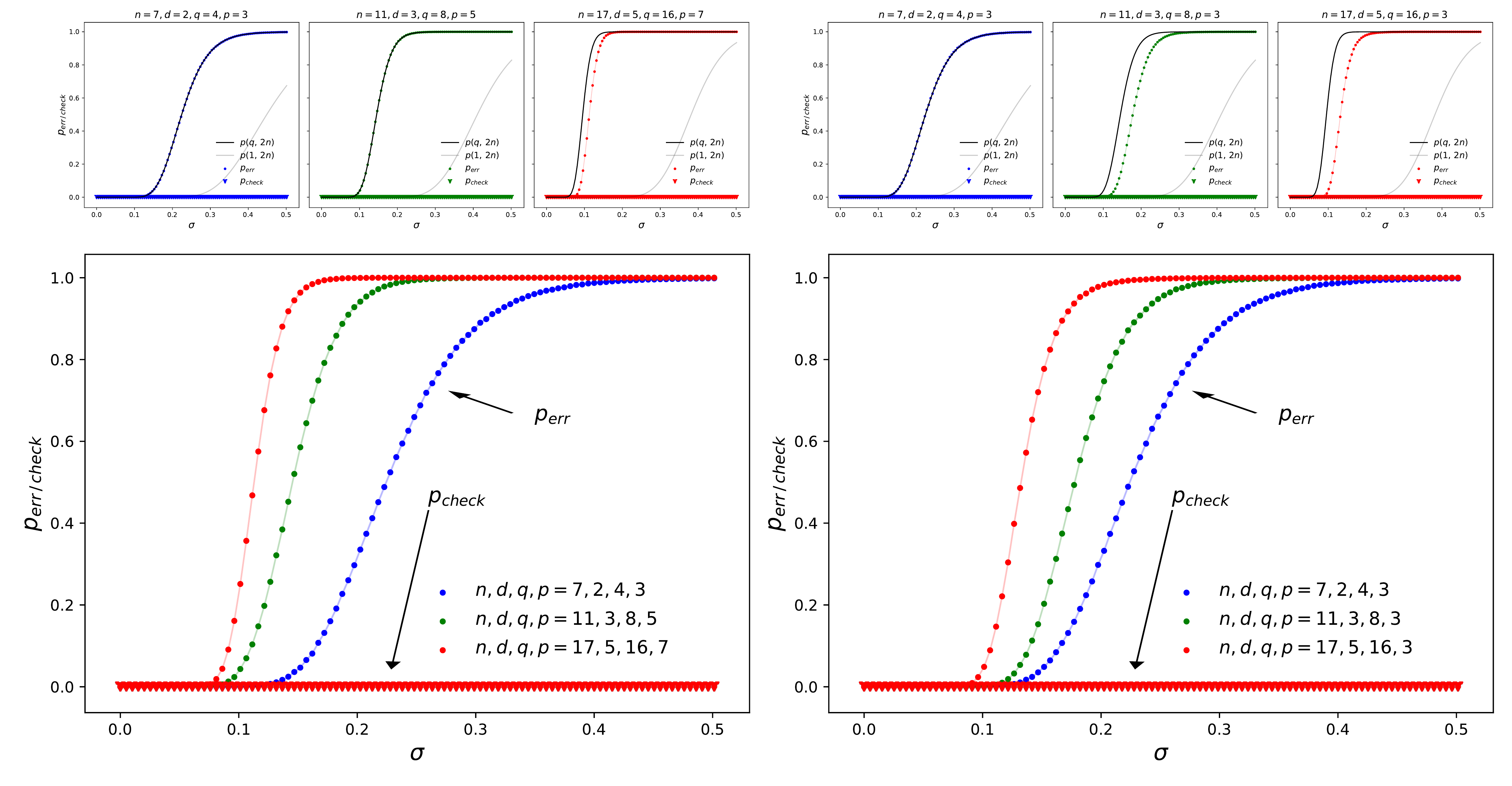}
 \caption{Numerical results for the NTRU-GKP codes using the NTRU decryption routine \texttt{NTRUDecode} for NTRU-GKP lattices where $h$ is invertible. (left) $p=3,5,7$ is running and (right) $p=3$ is fixed.}\label{fig:NTRU_dec}
 \end{figure}

Using these codes we simulate the error correction process on $N_{\rm samples}=10^5$ Gaussian distributed errors with physical variance $\sigma^2=2\pi\overline{\sigma}^2$.
The data displayed in 
fig.~\ref{fig:NTRU_dec_q8}  shows $\sigma^2=2\pi\overline{\sigma}^2$ on the x-axis denoting the \textit{physical} variance of Gaussian displacements and $p_{\rm err}$ as the logical error rate conditioned on successful decoding in the sense that the decoder successfully undid the syndrome.  We also plot $p_{\rm check}$,  denoting the rate of decoding failures,  i.e.,  the rate by which the decoder fails to output an error with the correct syndrome that is input to the decoder.  The standard deviation on the estimates for $p$ given by $\epsilon_{p}=\sqrt{p_{\rm err}(1-p_{\rm err})/N_{\rm sample}}$ is included in the plots but due to the sample number of $N_{\rm sample}=10^5$ is of negligible size. 
For comparison,  we also plot in black
\begin{align}
p(n,q,\sigma)&=1-\lrq{\int_{-\frac{\tilde{\Delta}}{2}}^{\frac{\tilde{\Delta}}{2}} \frac{e^{-\frac{x^2}{2\sigma^2}} }{\sqrt{2\pi\sigma^2}} dx}^{2n}\\&=1-\lrq{\erf\lr{\frac{\tilde{\Delta}}{2\sqrt{2}\sigma}}}^{2n},
\nonumber\\
\tilde{\Delta}&=\sqrt{\frac{2\pi}{2 q}},
\end{align}
which denotes the logical error probability of $n$ qudits with $d=2q$ encoded into the trivial sub-lattice $\CL_{\rm triv}=\sqrt{2q}\Z^{2n}$ corresponding to a hypercubic GKP code as well as in grey $p(n,1,\sigma)$, corresponding to the logical error probability of $n$ square GKP codes each encoding a single logical qubit. 
We further provide results for simulations of NTRU-GKP codes separately sampled from distributions $b)$  and $c)$ as denoted in 
fig.~\ref{fig:NTRU_sample}.
The parameters listed below in fig.~\ref{fig:params_a} reflect the codes simulated in 
figs.~\ref{fig:BAB_dec}, \ref{fig:BAB_dec_hinv},
\ref{fig:NTRU_dec}. 
In total, we make the following observations.

\begin{itemize}
    \item \texttt{BabaiDecode} has a rate of decoding failures matching $p(n,q,\sigma)$,  suggesting that decoding fails whenever the original error lies outside of the Voronoi Cell of $\CL_{\rm triv}^{\perp}=\Z /\sqrt{2 q}$.  When decoding is successful,  the logical error rate is negligible.
    
    \item \texttt{NTRUDecode} consistently corrects successfully,  i.e.,  returns the state to code space,  and has a conditional logical error rate that is smaller than $p(n,q,\sigma)$.  The logical error rate is however consistently larger than that of $n$ square GKP codes,  which is negligible in this parameter range. For $\sigma \approx 1.13$ we observe a ``threshold"-like behaviour in the transition between the sampled $n=17$ and $n=23$ code. 

    \item There appears to be no significant difference which $p$ we choose.
    
\end{itemize}

 The performance of these codes appears to be relatively poor when compared to more conventional multi-mode codes,  such as the \emph{toric-GKP code} \cite{toricGKP},  which we expect to be mainly the case due to the extremely small $n,q$ parameter regime we have simulated.
 
Further contributing factors to this observation may be that by decoding via essentially \texttt{MED} decoders,  we ignore a significant entropic contribution to the optimal \texttt{MLD} problem.  We have a number of minimal logical shifts $N_{\Delta^2} \geq n$ since the lattice $\CL$ is invariant under the cyclic shift $T:\, T^n=I$,  which is expected to be a relevant factor in the full \texttt{MLD} decoding problem.
 Another factor is that the NTRU decryption process used in \texttt{NTRUDecode} is in fact not tailored to a Gaussian distribution of random bits,  but rather is originally set up to decrypt a message hidden away using random strings $\bs{r}$ sampled from an uniform distribution.  \texttt{BabaiDecode} improves upon this fact in spirit by employing the \emph{nearest plane algorithm} in the decryption process,  but ignores the biasing of the error distribution,  eq.~\eqref{eq:res_prob},  from the first step of the decoding routine which is a necessary step in order to interpret the decryption process as a \texttt{CVP}.  
 
 It is interesting to observe that \texttt{BabaiDecode} consistently displays a negligible logical error rate but quickly rises to high decoding failure rate,  which worsens as the code is scaled up and that for \texttt{NTRUDecode} we do observe a parameter range where the decoder displays a lower logical error rate than $p(n,q,\sigma)$.  This shows that the decoder indeed non-trivially decodes errors.
Overall,  it appears necessary to perform larger scale numerical studies at large $n, q$ to examine the possibility of a threshold.
 The \texttt{sagemath} \cite{sagemath} and \texttt{python} code as well as all numerical data presented here is publicly available under
 ref.~\cite{GitLink}. \texttt{sagemath} functionalities to construct NTRU lattices are partially adapted from ref.~\cite{LatticeHacks}.

\begin{figure}
\centering
        \begin{minipage}{0.4\textwidth}
         \begin{tabular}{c|c|c}
           $n,d,q,p$    &  $\lambda_1\lr{L_{\rm cs}}$ & $\Delta$ \\\hline
           $7,2,4,3$    & $2.65$ & $ 0.94$ \\
           $11,3,8,5$   & $3.32$ & $ 0.83$\\
           $17,5,16,7$  & $4.12$ & $ 0.74$ \vspace{.2cm}\\
           $7,2,4,3$    & $2.65$ & $ 0.94$\\
           $11,3,8,3$   & $3.32$ & $0.83$\\
           $17,5,16,3$  & $4.12$ & $0.74$
        \end{tabular}
        \end{minipage}\hspace{.2cm}%
        \begin{minipage}{0.4\textwidth}
         \begin{tabular}{c|c|c}
           $n,d,q,p$    &  $\lambda_1\lr{L_{\rm cs}}$ & $\Delta$ \\\hline
           $7,2,4,3$    & $2.83$ & $ 1$\\
           $11,3,8,5$   & $4.24$ & $ 1.06$ \\
           $17,5,16,7$  & $7.07$ & $ 1.24$ \vspace{.2cm}\\ 
           $7,2,4,3$    & $2.83$ & $ 1$ \\
           $11,3,8,3$   & $4.24$ & $1.06$\\
           $17,5,16,3$  & $6.93$ & $1.23$
        \end{tabular}
        \end{minipage}%
        \caption{Lattice parameters for random NTRU lattices. The table on the right summarizes the results when additionally $h$ is required to be invertible.}\label{fig:params_a}
\end{figure}

\end{document}